\documentclass[journal]{IEEEtran}
\def\*#1{\boldsymbol{#1}}

\ifCLASSINFOpdf
\else
\fi
\usepackage{amsmath,amssymb,graphicx,cases}
\usepackage{url,hyperref}
\begin{document}
\title{Reconstruction Error Bounds for Compressed Sensing under Poisson or Poisson-Gaussian Noise Using Variance Stabilization Transforms}

\author{Deepak~Garg,~\IEEEmembership{Student Member,~IEEE,} 
Pakshal~Bohra, Karthik S. Gurumoorthy, 
        and~Ajit~Rajwade,~\IEEEmembership{Member,~IEEE}
% <-this % stops a space
\thanks{Deepak Garg and Pakshal Bohra are both first authors with equal contribution. Deepak Garg and Ajit Rajwade are with the Department of Computer Science and Engineering at IIT Bombay. Pakshal Bohra is with the department of Electrical Engineering at IIT Bombay. Karthik S. Gurumoorthy is with the International Center for Theoretical Sciences. Their email addresses are \url{19deepak94@gmail.com,pakshalbohra@gmail.com, karthik.gurumoorthy@icts.res.in, ajitvr@cse.iitb.ac.in}. Corresponding author is AR. AR acknowledges support from IITB Seed Grant 14IRCCSG012. KSG acknowledges the support of the AIRBUS Group Corporate Foundation Chair in Mathematics of Complex Systems established in ICTS-TIFR.}% <-this % stops a space
}


\markboth{Submitted}%
{Patil \MakeLowercase{\textit{et al.}}: Bare Demo of IEEEtran.cls for IEEE Journals}



\maketitle

\begin{abstract}
Most existing bounds for signal reconstruction from compressive measurements make the assumption of additive signal-independent noise. However in many compressive imaging systems, the noise statistics are more accurately represented by Poisson or Poisson-Gaussian noise models. In this paper, we derive upper bounds for signal reconstruction error from compressive measurements which are corrupted by Poisson or Poisson-Gaussian noise. The features of our bounds are as follows: (1) The bounds are derived for a computationally tractable convex estimator with statistically motivated parameter selection. The estimator penalizes signal sparsity subject to a constraint that imposes a novel statistically motivated upper bound on a term based on variance stabilization transforms to approximate the Poisson or Poisson-Gaussian distributions by distributions with (nearly) constant variance. (2) The bounds are applicable to signals that are sparse as well as compressible in any orthonormal basis, and are derived for compressive systems obeying realistic constraints such as non-negativity and flux-preservation. We present extensive numerical results for signal reconstruction under varying number of measurements and varying signal intensity levels. Ours is the first piece of work to derive bounds on compressive inversion for the Poisson-Gaussian noise model.
\end{abstract}


% \fi
\IEEEpeerreviewmaketitle

\section{Introduction}
\IEEEPARstart{C}{ompressed} sensing (CS) is a flourishing branch of signal processing with many theoretical and algorithmic advances, along with emerging applications in the form of actual systems in medicine, astronomy, photography and various other fields. Theoretical bounds for performance of compressive reconstruction algorithms have shown great promise \cite{Candes2008}, but most of them are based on the assumption of additive signal independent noise. However the noise in many compressive imaging systems can be more accurately described as Poisson-Gaussian. The Poisson component, which is signal dependent, is typically known to emerge from photon-counting principles in the acquisition of signals. The Gaussian component is signal-independent and is due to fluctuations in the electronic parts of the imaging system. The Poisson component is quite dominant particularly at lower signal intensities \cite{Trussell2012}, and is a non-additive form of noise. Given a non-negative signal $\boldsymbol{x} \in \mathbb{R}^m$ and a compressive measuring device with a non-negative sensing matrix $\boldsymbol{\Phi} \in \mathbb{R}^{N \times m}, N \ll m$, the measurement vector $\boldsymbol{y} \in \mathbb{R}^m$ can be described as follows:
\begin{equation}
\boldsymbol{y} \sim \alpha \textrm{Poisson}(\boldsymbol{\Phi x}) + \boldsymbol{\eta}, \boldsymbol{\eta} \sim \mathcal{N}(g,\sigma^2),
\end{equation}
where $\alpha$ represents a gain factor, and $g,\sigma$ represent the mean and standard deviation of the Gaussian component respectively. The Gaussian component of the noise cannot be ignored, and such a mixed Poisson-Gaussian noise model is ubiquitous in imaging systems in astronomy \cite{Murtagh1995}, microscopy \cite{Delpretti2008} and compressive imagers such as the Rice Single Pixel camera \cite{DuBosq2016,Duarte2008}, to name a few.

There exists a large amount of literature on denoising of signals or images under Poisson-Gaussian noise. For instance, recent work in \cite{Chouzenoux2015} denoises and deblurs images using an exact Poisson-Gaussian likelihood, which is approximated in a very principled way during an iterative optimization. Earlier work on image denoising using this model includes approximations based on variance stabilization transforms \cite{Murtagh1995} or PURElet-based approaches \cite{Luisier2011}, among others. However, this noise model has not been presented heretofore in the context of CS, and in particular with a derivation of performance bounds. There does exist fairly recent literature on performance bounds for CS under purely Poisson noise using either the penalized Poisson negative log-likelihood or the LASSO (see Section \ref{sec:discussion} for a detailed discussion), or using least squares estimation for Poisson inverse problems with $N > m$ \cite{Shin2016}. Efficient algorithms have also been proposed for Poisson CS \cite{Harmany2012,Lingenfelter2009, Starck2010, Zhang2008} or Poisson deconvolution \cite{Dupe2009}. A comprehensive survey of algorithms and applications of Poisson inverse problems has been presented in \cite{Hohage2016}.

In this paper, we derive performance bounds for CS under Poisson noise using a variance stabilization transform (VST) approach. As has been shown in \cite{Anscombe1948}, if $y \sim \textrm{Poisson}(\lambda)$, then $\sqrt{y + \frac{3}{8}}$ has variance approximately $\frac{1}{4}$ and mean $\sqrt{\lambda+\frac{3}{8}}$ when $\lambda \rightarrow \infty$. This motivates the following objective function for compressive inference:
\begin{equation}
\textrm{min} \|\boldsymbol{\theta}\|_1 \textrm{ subject to } \|\sqrt{\boldsymbol{y}+c}-\sqrt{\boldsymbol{\Phi \Psi \theta}+c}\|_2 \leq \varepsilon, \boldsymbol{\Psi \theta} \succeq \boldsymbol{0}
\end{equation}
where $\boldsymbol{\Psi}$ is a $m \times m$ orthonormal basis in which the signal $\boldsymbol{x}$ yields a sparse set of coefficients $\boldsymbol{\theta} = \boldsymbol{\Psi}^T \boldsymbol{x}$, $c$ is a coefficient that defines the VST (\textit{e.g.}, $c = \frac{3}{8}$ for the Anscombe transform) and the symbol $\succeq$ in $\* a \succeq \*b$ means that $a_i \geq b_i$ for every index $i$ in vectors $\*a$ and $\*b$. Here $\varepsilon$ is a statistically motivated upper bound on $\|\sqrt{\boldsymbol{y}+c}-\sqrt{\boldsymbol{\Phi \Psi \theta}+c}\|_2$ where the noise term $\sqrt{\boldsymbol{y}+c}-\sqrt{\boldsymbol{\Phi \Psi \theta}+c}$ has variance approximately $\frac{1}{4}$ (\emph{after} application of the VST to the noisy CS measurements). We also extend these bounds to the case of Poisson-Gaussian noise.

The contribution of our work is summarized as follows:
\begin{enumerate}
\item To the best of our knowledge, this is the first piece of work to provide performance bounds for CS under Poisson-Gaussian noise. In fact, we have a unified approach to handle Poisson as well as Poisson-Gaussian noise.
\item Our bounds apply to a computationally tractable and probabilistically motivated estimator, under realistic CS matrices, and for sparse or compressible signals in any orthonormal basis. A detailed comparison with earlier work is presented in Section \ref{sec:discussion}.
\item Due to the VST, our estimator allows for very principled, statistically motivated parameter tuning, since the term $\|\sqrt{\boldsymbol{y}+c}-\sqrt{\boldsymbol{\Phi \Psi \theta}+c}\|^2_2$ is a metric and since (as we show later in the paper) the magnitude of the difference term, \textit{i.e.} $\|\sqrt{\boldsymbol{y}+c}-\sqrt{\boldsymbol{\Phi \Psi \theta}+c}\|_2$, has a bounded variance which does not depend on the original signal or the number of measurements. This statistically motivated parameter tuning is different from the case of the Poisson negative log-likelihood which is not a metric, which does not have a signal-independent value, and where choosing the regularization parameter for signal sparsity is not easy in practice. Again, see Section \ref{subsec:results_P} and \ref{sec:discussion}.
\end{enumerate}
A part of this work earlier appeared in our conference paper \cite{Garg2017}, but this work contains an extension to the Poisson-Gaussian case, as well as many refinements to the theory and experiments for the Poisson noise case.

This paper is organized as follows. Some preliminaries are presented in Section \ref{sec:prelims}, the main theoretical results are derived in \ref{sec:theory} along with a discussion, numerical results are presented in Section \ref{sec:results}, followed by a summary of the contributions, a more detailed comparison with existing work and directions for future work in Section \ref{sec:discussion}.

\section{Preliminaries}
\label{sec:prelims}
In this section, we go over some preliminary concepts briefly, so as to make the paper self-contained. 
\subsection{Construction of Sensing Matrices}
\label{subsec:sm}
We construct a sensing matrix $\boldsymbol{\Phi}$ that corresponds to the forward model of a real optical system, based on the approach in \cite{Raginsky2010}. Clearly $\boldsymbol{\Phi}$ has to satisfy certain constraints natural to a realizable imaging system - non-negativity and flux preservation. The latter is due to the fact that the total photon-count of the noise-free measurement $\boldsymbol{\Phi x}$ can never exceed that of the original signal $\boldsymbol{x}$, \textit{i.e.}, $\sum_{i=1}^N (\boldsymbol{\Phi x})_i \leq \sum_{k=1}^m x_k$. This in turn imposes the constraint that every column of $\boldsymbol{\Phi}$ must sum up to a value no more than 1, i.e. $\forall j, \sum_{i=1}^N {\Phi}_{ij} \leq 1$. 

One major difference between Poisson CS and conventional CS emerges from the fact that conventional randomly generated sensing matrices which obey restricted isometry (RIP) do not follow the aforementioned physical constraints. This is a drawback as the RIP is a well-known sufficient condition which guarantees bounds on compressive recovery. We now construct a sensing matrix $\boldsymbol{\Phi}$ which has only zero or (scaled) ones as entries. Let us define $p$ to be the probability that a matrix entry is 0, then $1-p$ is the probability that the matrix entry is a scaled 1. Let $\boldsymbol{Z}$ be a $N \times m$ matrix whose entries $Z_{i,j}$ are i.i.d random variables taking only these two different values, \textit{i.e.},
\begin{subnumcases}{Z_{i,j}=}
-\sqrt{\dfrac{1-p}{p}} & with probability $p$,
\\
\sqrt{\dfrac{p}{1-p}} & with probability $1-p$.
\end{subnumcases}
Let us define $\boldsymbol{\tilde{\Phi}} \triangleq \dfrac{\boldsymbol{Z}}{\sqrt{N}}$. For $p = 1/2$, the matrix $\boldsymbol{\tilde{\Phi}}$ now follows RIP of order $2s$ with a very high probability given as $1-2e^{-Nc(1+\delta_{2s})}$ where $\delta_{2s}$ is its RIC of order $2s$ and function $c(h) \triangleq \dfrac{h^2}{4}-\dfrac{h^3}{6}$ \cite{Baraniuk2008}. In other words, for any $2s$-sparse signal $\boldsymbol{\rho}$, the following holds with high probability
\begin{equation*} (1-\delta_{2s})\|\boldsymbol{\rho}\|^2_2 \leq \|\boldsymbol{\tilde{\Phi}\rho}\|^2_2 \leq (1+\delta_{2s})\|\boldsymbol{\rho}\|^2_2. \end{equation*} 
Given any orthonormal matrix $\boldsymbol{\Psi}$, arguments in \cite{Baraniuk2008} show that $\boldsymbol{\tilde{\Phi}\Psi}$ also obeys the RIP of the same order as $\boldsymbol{\tilde{\Phi}}$. 

However $\boldsymbol{\tilde{\Phi}}$ will clearly contain negative entries with very high probability, which violates the constraints of a physically realizable system. To deal with this, we can construct the flux-preserving and non-negative sensing matrix $\boldsymbol{\Phi}$ from $\boldsymbol{\tilde{\Phi}}$ as follows \cite{Raginsky2010}:
\begin{equation} 
\boldsymbol{\Phi} = \sqrt{\dfrac{p(1-p)}{N}} \boldsymbol{\tilde{\Phi}} + \dfrac{(1-p)}{N}\boldsymbol{1}_{N \times m}, 
\label{eq:Phi} 
\end{equation}
which ensures that each entry of $\boldsymbol{\Phi}$ is either $0$ or $\dfrac{1}{N}$. One can easily check that $\boldsymbol{\Phi}$ satisfies both the non-negativity as well as flux-preservation properties. 

\subsection{Variance Stabilization Transforms}
VSTs are a popular method of converting Poisson data into data that are approximately Gaussian. In particular, \cite{Anscombe1948} proves that if $y \sim \textrm{Poisson}(\lambda)$, then we have the following:
\begin{eqnarray}
E(\sqrt{y+c}) = \sqrt{\lambda+c} - \frac{1}{8\sqrt{\lambda}} + \mathcal{O}(\lambda^{-1.5}) \\
\textrm{Var}(\sqrt{y+c}) = \frac{1}{4} + \frac{3-8c}{32\lambda} + \mathcal{O}(\lambda^{-2}).
\end{eqnarray}
Setting $c = \frac{3}{8}$ yields the so-called Anscombe Transform (AT) and produces data with a `stable' noise variance of approximately $\frac{1}{4}$ and a mean of approximately $\sqrt{\lambda + c}$. The higher order moments are approximately zero for a reasonably large $\lambda$. The approximation to the mean is further approximated as $\sqrt{\lambda}$ in some papers \cite{Hohage2016}. All these approximations improve as $\lambda$ grows beyond 4, and the noise distribution becomes closer and closer to $\mathcal{N}(0,\frac{1}{4})$ as shown rigorously in \cite{Curtiss1943}. In the case of Poisson-Gaussian noise, i.e. when $y \sim \alpha \textrm{Poisson}(\lambda) + \eta$ where $\eta \sim \mathcal{N}(g,\sigma^2)$, the AT is replaced by the Generalized AT (GAT) which is given as $t = \frac{1}{\alpha}\sqrt{\alpha y + \frac{3}{8}\alpha^2 + \sigma^2 - \alpha g}$. As $\lambda$ grows in value, it can be shown \cite{Murtagh1995} that $t$ has a mean of $\sqrt{\lambda + \frac{3}{8}\alpha + \frac{\sigma^2 - \alpha g}{\alpha}}$ and variance of approximately $\frac{1}{4}$. In this paper, we keep $\alpha = 1, g = 0$ for simplicity, although our framework is general enough to handle deviations from this assumption.

\section{Theory}
\label{sec:theory}
The main theoretical development is presented in this section. First, for noisy measurements $\boldsymbol{y} \sim \textrm{Poisson}(\boldsymbol{\Phi x})$, we prove that the quantity $R(\boldsymbol{y},\boldsymbol{\Phi x}) \triangleq \|\sqrt{\boldsymbol{y}+c}-\sqrt{\boldsymbol{\Phi x}+c}\|_2$ (henceforth called the `residual magnitude') has a mean which is $\mathcal{O}(\sqrt{N})$ and a variance which is constant (independent of the signal $\boldsymbol{x}$ and also suprisingly independent of the number of measurements $N$) as long as $\boldsymbol{\Phi x} \succeq \boldsymbol{1}$. This result is extended to the case of Poisson-Gaussian noise. Using these results, we then state and prove two theorems for upper error bounds for the reconstruction of a signal from Poisson corrupted CS measurements in a realistic system as per Eqn. \ref{eq:Phi}. Another two theorems are stated and proved for the case of Poisson-Gaussian CS. An extensive discussion on the theorem statements is presented. The proofs of the theorems on error bounds follow the broad technique from \cite{Candes2008}.

\subsection{Theorem for Properties of the Residual Magnitude}
\begin{figure*}[!t]
\centering
\includegraphics[width=1.75in]{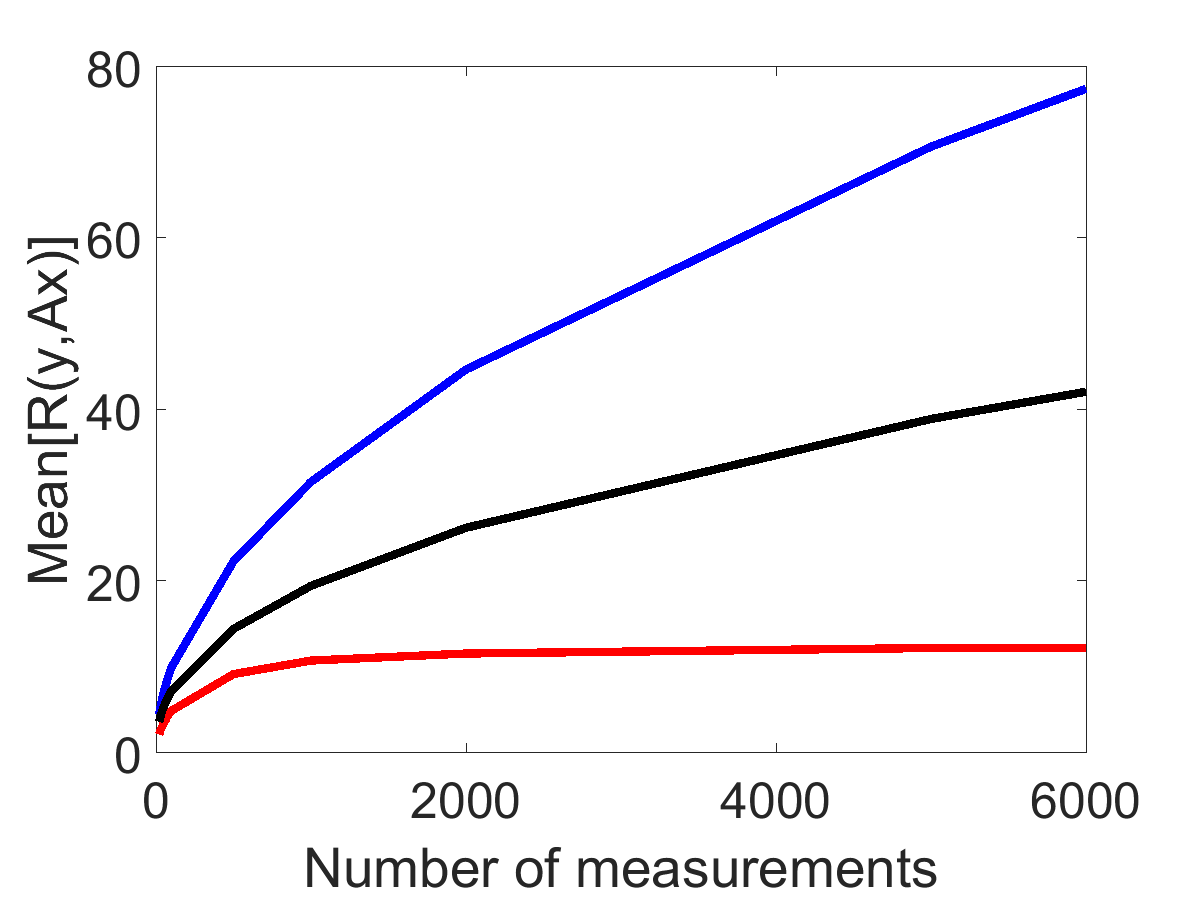}
\includegraphics[width=1.75in]{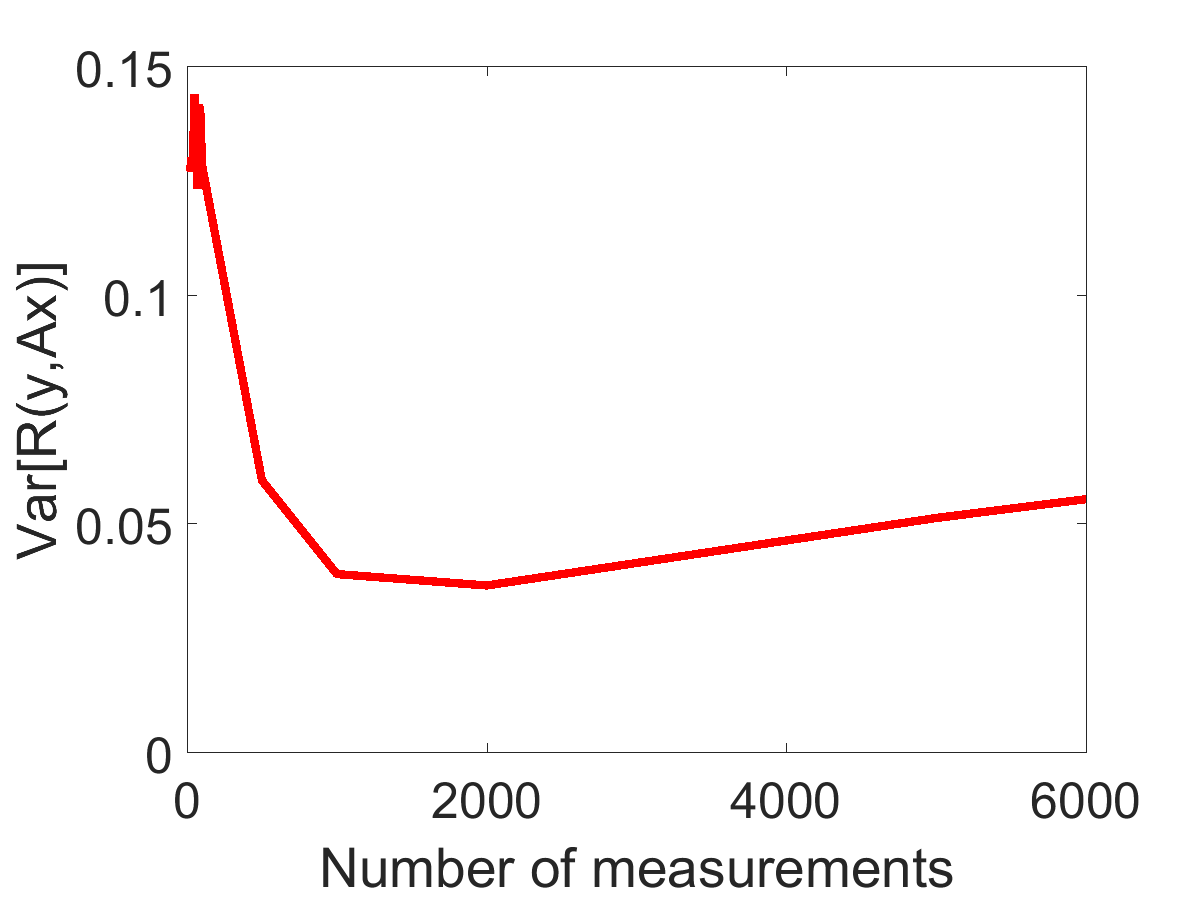}
\includegraphics[width=1.75in]{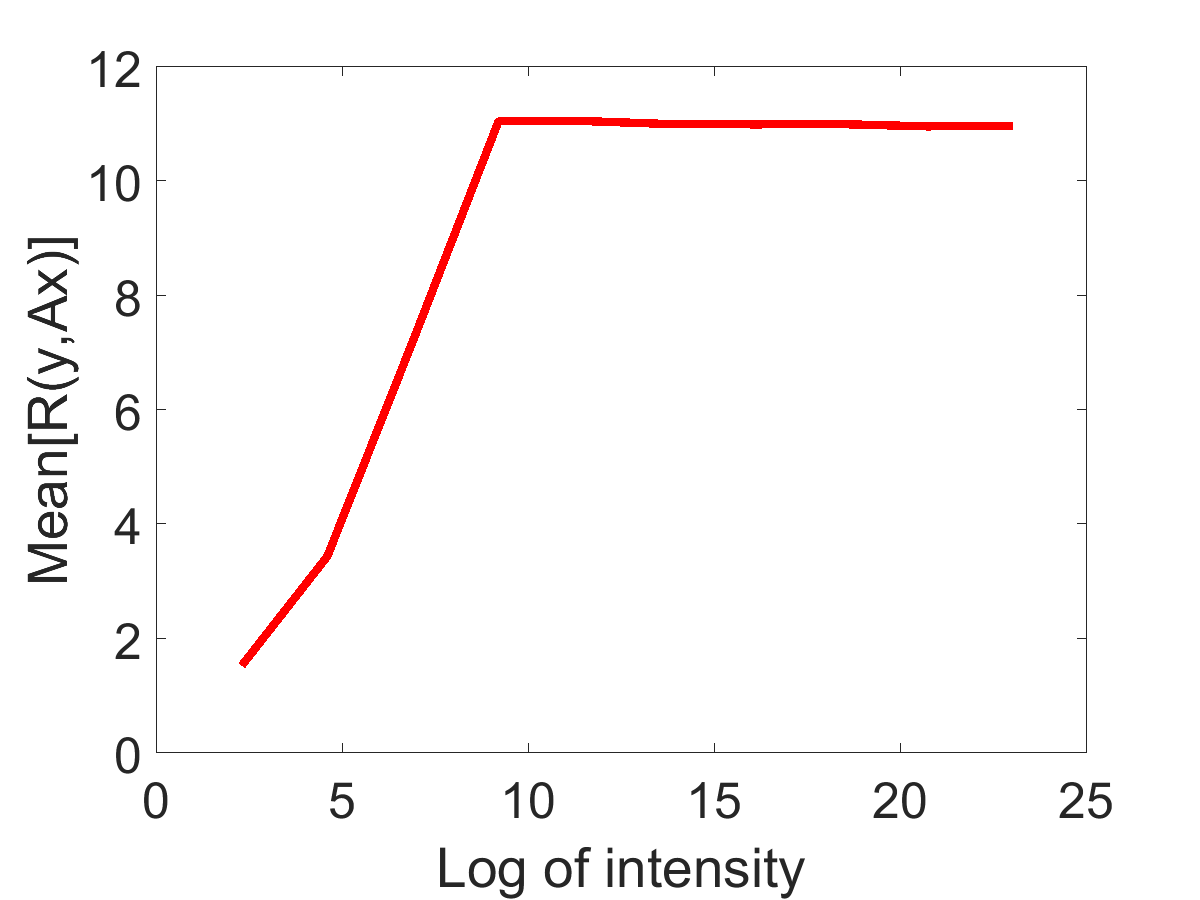}
\includegraphics[width=1.75in]{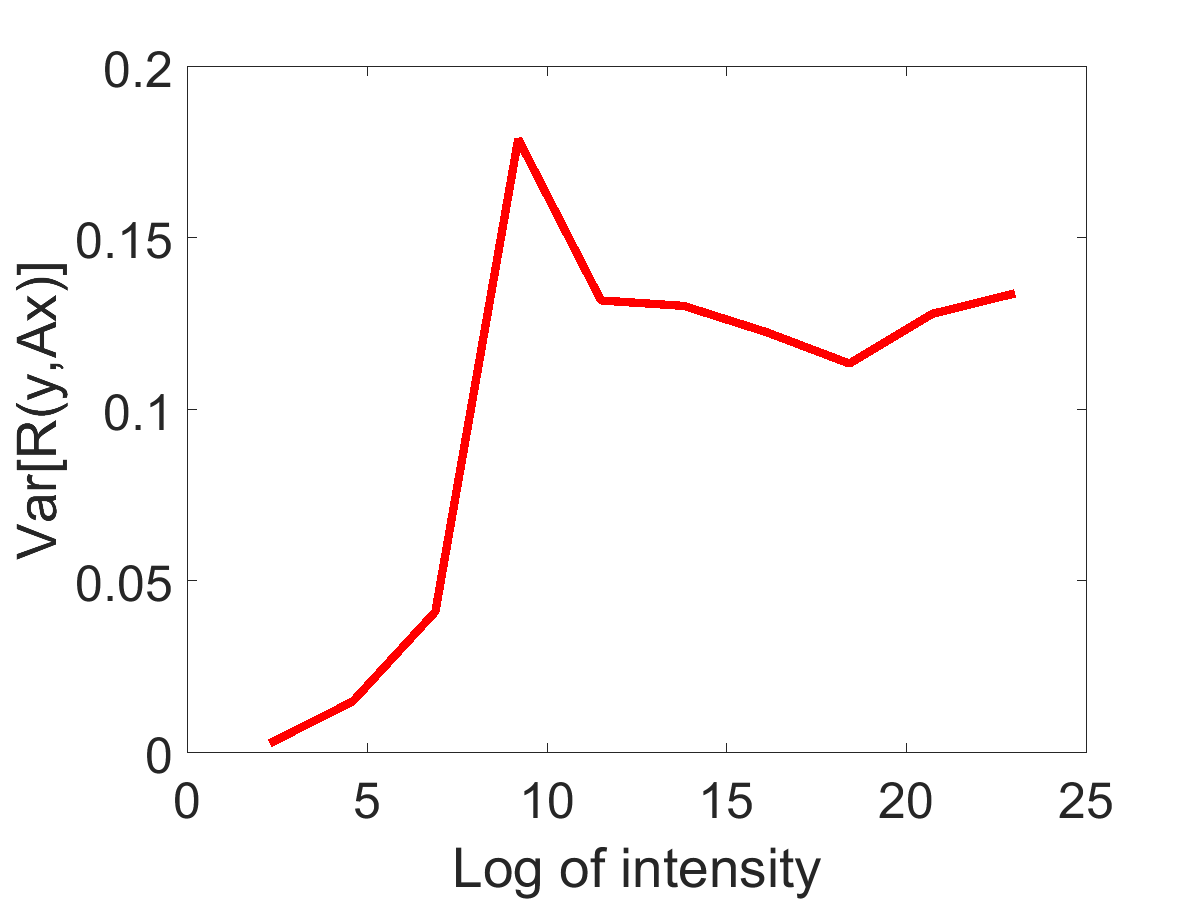}
\includegraphics[width=1.75in]{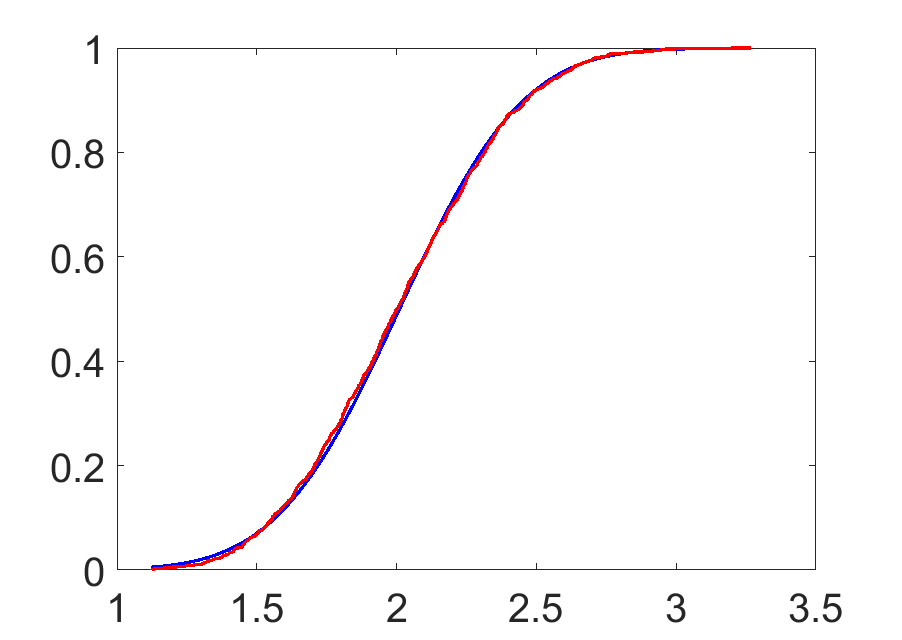} 
\caption{In the left to right, top to bottom order. First two sub-figures: Plot of mean and variance of the values of $R(\boldsymbol{y},\boldsymbol{\Phi x})$ versus $N$ for a fixed $I = 10^3$ for a signal of dimension $m = 1000$. (For the leftmost sub-figure in the first row, the blue line represents the plot of $\sqrt{N}$ and the black line represents $N^{0.43}$.) Third and fourth sub-figures: Plot of mean and variance of the values of $R(\boldsymbol{y},\boldsymbol{\Phi x})$ versus $\log(I)$ for a fixed $N=500$ for a signal of dimension $m = 1000$. Last sub-figure: Empirical CDF of $R(\boldsymbol{y},\boldsymbol{\Phi x})$ (red curve) for $N = 20, I = 10^3, m = 1000$ compared to a Gaussian CDF (blue curve) with mean and variance equal to that of the values of $R(\boldsymbol{y},\boldsymbol{\Phi x})$. The curves overlap significantly as the empirical CDFs are very close. Scripts for reproducing these results are available at \cite{suppcode}.}
\label{fig:anscombe_mean_var}
\end{figure*}

The theorem we present in this section was inspired by our simulations with the quantity $R(\boldsymbol{y},\boldsymbol{\Phi x})$ defined above. We simulated Poisson-corrupted CS measurements $\*y \sim \*\Phi \*x$ for sensing matrix $\* \Phi \in \mathbb{R}^{N \times m}$ as per Eqn. \ref{eq:Phi} and for a non-negative signal $\boldsymbol{x}$ of $m = 1000$ dimensions. The signal intensity was $I = 1000$. The signal values were generated from $\textrm{Unif}[0,1]$. The chosen values of $N$ were from 20 to 6000. For each $N$, 2000 measurements were generated keeping $\* \Phi, \*x$ fixed. We empirically observed that $E[R(\boldsymbol{y},\boldsymbol{\Phi x})]$ was $\mathcal{O}(\sqrt{N})$, i.e. independent of $I$. We also observed that $\textrm{Var}[R(\boldsymbol{y},\boldsymbol{\Phi x})]$ was upper bounded by a small constant value around 0.14 independent of both $I$ and $N$. We repeated this experiment for a fixed $N = 500$ and fixed $\boldsymbol{x}/\|\boldsymbol{x}\|_1$  but varying $I$ from $10^2$ to $10^9$ in powers of 10. Again, we observed the same properties of $E[R(\boldsymbol{y},\boldsymbol{\Phi x})]$ and $\textrm{Var}[R(\boldsymbol{y},\boldsymbol{\Phi x})]$. Moreover, we observed that the empirical CDF of the values of $R(\boldsymbol{y},\boldsymbol{\Phi x})$ was similar to a Gaussian. These results are shown in Fig. \ref{fig:anscombe_mean_var}. These results were independent of the specific instances of $\*x, \*y, \*\Phi$.

\textbf{Theorem 1:} Let $\boldsymbol{y} \in \mathbb{Z}^N_{+}$ be a vector of independent CS measurements such that $y_i \sim \textrm{Poisson}[(\boldsymbol{\Phi x})_i]$ where $\boldsymbol{\Phi} \in \mathbb{R}^{N \times m}$ is a non-negative flux-preserving matrix as per Eqn. \ref{eq:Phi} and $\boldsymbol{x} \in \mathbb{R}^m$ is a non-negative signal. Define $\gamma_i \triangleq (\boldsymbol{\Phi x})_i$. Then we have:
\begin{enumerate}
\item $E[R(\boldsymbol{y},\boldsymbol{\Phi x})] \leq \sqrt{N/2}$
\item Define $v \triangleq \textrm{Var}[R(\boldsymbol{y},\boldsymbol{\Phi x})]$. Then we have \\
$v \leq \dfrac{\sum_{i=1}^N \frac{\gamma_i(1+3\gamma_i)}{(\gamma_i+c)^2}} {\sum_{i=1}^N \textrm{max}(0,\frac{\gamma_i}{4(\gamma_i+c)}-\frac{\gamma_i}{8(\gamma_i+c)^2})}$
\item If $\forall i, \gamma_i \geq 1$, then $v \leq \frac{3N/4+N/4}{N(2c+1)/(8(1+c)^2))} \lessapprox 10.85$
\item $P\Big(R(\* y, \*\Phi\*x) \leq \sqrt{N}(\frac{1}{\sqrt{2}}+3.29)\Big) \geq 1-1/N$.
\end{enumerate}
All statements of this theorem are proved in Section \ref{subsec:thm1}. We make a few comments below:
\begin{enumerate}
\item $E[R(\boldsymbol{y},\boldsymbol{\Phi x})]$ \emph{does not} increase with $I$. This property is \emph{not} shared by the negative log-likelihood of the Poisson distribution. 
\item The third statement is clearly a corollary to the second statement, i.e. the variance bound. In practice, we have observed a smaller value of this constant close to 0.14 \emph{even when} the condition that $\forall i, \gamma_i \geq 1$ is violated, even though the predicted upper bound on the variance is larger. The assumption that $\gamma_i \geq 1$, is not restrictive in most signal or image processing applications, except those that work with extremely low intensity levels. But in such cases the performance of Poisson CS is itself very poor due to the very low SNR \cite{Jiang2015}. 
\item The last statement of this theorem can be further tightened to yield a probability of $1-2e^{-N/2}$ by using the central limit theorem (CLT). Of course, the latter is an asymptotic result and hence for a finite value of $N$, it is an approximation. However, the approximation is empirically observed to be tight even for small $N \sim 20$ as confirmed by a Kolmogorov-Smirnov test even at 1\% significance (see \cite{suppcode}). Further details can be found at the end of the proof in Section \ref{subsec:thm1}.
\item The bounds in this theorem do \emph{not} assume (or require) that $\sqrt{\boldsymbol{y}+c}-\sqrt{\boldsymbol{\Phi x}+c}$ is Gaussian distributed. Indeed such an assumption would not be rigorous enough. This is because as shown in \cite{Curtiss1943}, the Gaussianity is obeyed only asymptotically when the mean of $\*y$ tends to infinity. 
\end{enumerate}

\subsection{Key Theorem for Poisson CS}
\textbf{Theorem 2 : } Consider a non-negative signal $\boldsymbol{x}$ with total intensity $I \triangleq \|\boldsymbol{x}\|_1$ expressed using the orthornormal basis $\boldsymbol{\Psi}$ in the form $\boldsymbol{x} = \boldsymbol{\Psi\theta}$. Consider Poisson corrupted CS measurements of the form $\boldsymbol{y} \sim \textrm{Poisson}(\boldsymbol{\Phi x})$ where $\boldsymbol{\Phi}$ is constructed as per Eqn. \ref{eq:Phi}. Define $\*A \triangleq \*\Phi \*\Psi$ so that $\*\Phi \*x = \*A \*\theta$.  Let $\boldsymbol{\theta^\star}$ be the result of the following optimization problem:
\begin{eqnarray}
(P1): \textrm{min} \|\boldsymbol{\theta}\|_1 \textrm{ such that } \|\sqrt{\boldsymbol{y}+c}-\sqrt{\boldsymbol{A \theta}+c}\|_2 \leq \varepsilon, \\ \nonumber
\|\boldsymbol{\Psi \theta}\|_1 = I, \boldsymbol{\Psi \theta} \succeq \boldsymbol{0},
\end{eqnarray}
where $\varepsilon \triangleq \sqrt{N}(3.29+1/\sqrt{2})$ is a statistical upper bound (that holds with a high probability $1-1/N$) on the magnitude of the noise in the measurements \emph{after} application of the AT. Let $\boldsymbol{\theta_s}$ denote a vector containing the $s$ largest magnitude elements of $\boldsymbol{\theta}$ with the rest being 0. If $\boldsymbol{\widetilde \Phi}$ obeys RIP of order $2s$ with RIC $\delta_{2s} < \sqrt{2}-1$, and the condition $\boldsymbol{\Phi x} \succeq \boldsymbol{1}$ holds, then we have for any $\kappa > 0$:
\begin{eqnarray}
%E \bigg( \frac{\|\boldsymbol{\theta}-\boldsymbol{\theta^\star}\|_2}{I} \bigg) \leq C_1\sqrt[]{\dfrac{N}{2}} \sqrt{\frac{1}{I}+\frac{cN}{I^2}} + \frac{C_2s^{-\frac{1}{2}}\|\boldsymbol{\theta}-\boldsymbol{\theta_s}\|_1}{I},\\
P \bigg( \frac{\|\boldsymbol{\theta}-\boldsymbol{\theta^\star}\|_2}{I} \leq C_1\sqrt[]{N}\tau \sqrt{\frac{1}{I}+\frac{cN}{I^2}} + \frac{C_2s^{-\frac{1}{2}}\|\boldsymbol{\theta}-\boldsymbol{\theta_s}\|_1}{I} \bigg) \\ \nonumber
\geq 1-\kappa^2/N \textrm{ where } \tau \triangleq (3.29/\kappa+1/\sqrt{2}).
%P \bigg( \frac{\|\boldsymbol{\theta}-\boldsymbol{\theta^\star}\|_2}{I} \leq C_1\sqrt[]{N \tilde{\tau}} \sqrt{\frac{1}{I}+\frac{cN}{I^2}} + \frac{C_2s^{-\frac{1}{2}}\|\boldsymbol{\theta}-\boldsymbol{\theta_s}\|_1}{I} \bigg) \\ \nonumber
%\geq 1-e^{-N\tau} \textrm{ where } \tilde{\tau} \triangleq (1+2\tau+\sqrt{2\tau}) \textrm{ for some } \tau > 0.
\end{eqnarray}
This theorem is proved in Section \ref{subsec:thm2}. Comments on this theorem follow.\\\\
\textbf{Remarks on the Theorem and its Proof:}
\begin{enumerate}
\item The tighest upper bounds we have are for $c = 0$, i.e. the original square-root VST developed by Bartlett \cite{Bartlett1936}.
\item Our proof architecture is inspired from \cite{Candes2008}, but the points of departure are steps 2(a), 2(b), 2(c) as well as step 4(a) which gives a relationship between $\|\boldsymbol{Ah}\|_2$ and $\|\boldsymbol{Bh}\|_2$. These steps exploit the non-negativity and flux-preserving property of $\boldsymbol{\Phi}$, and the constraint $\|\boldsymbol{\Psi \theta}\|_1 = I$. See Section \ref{subsec:thm2}.
\item Given that we are dealing with a \emph{Poisson} inverse problem, it is more intuitive to analyze the \emph{relative} reconstruction error (RRE) rather than the (absolute) reconstruction error. This is because as the mean of the Poisson distribution increases, so does its variance, causing an increase in the mean squared error but a decrease in the relative mean squared error. 
\item Notice that our derived RRE bound is inversely proportional to the signal intensity $I$. For a fixed $I$, if $N$ is increased, the incident photon flux $I$ is distributed across the $N$ measurements, causing a decrease in SNR per measurement and possibly degrading performance. In fact, this affects the bounds in the $c \neq 0$ case. This phenomenon differs from CS under Gaussian noise, and has earlier been noted in \cite{Raginsky2010,YXie2013,Cao2016}. For $c = 0$, however, the flux-preserving nature of the matrix does not affect the bounds, rather the $\sqrt{N}$ term is due to the fact that the variance of the noise after VST is a constant independent of $N$ although there are $N$ measurements. This is similar to Equation (17) of \cite{Candes2006} for pure Gaussian noise.
\item As $s$ increases, the restricted isometry constant (RIC) $\delta_{2s}$ of the sensing matrix will increase. Hence the constants $C_1$ and $C_2$ will increase since they are monotonically increasing functions of $\delta_{2s}$. Hence as per the bounds we have derived, the upper bounds on the performance will actually increase with $s$. In fact, for a fixed number of measurements $N$, an increase in $s$ may cause the sensing matrix to no longer obey the restricted isometry property (RIP). This phenomenon directly follows \cite{Candes2008} and is not exclusive to the technique and bounds developed by us. 
\item Our experimental results in the next section show that knowledge of $I$ is not necessary, although we required it for our theoretical analysis.
\item The RRE bounds are also applicable to the Freeman-Tukey transform \cite{Freeman1950} given as $\sqrt{y} + \sqrt{y+1}$ with minor changes to the constant $C_1$.
\item As has been mentioned earlier, the VST approximation is not so accurate for measurements with low mean, however at such low intensity levels Poisson CS is considered to be undesirable in itself \cite{Jiang2015}.
\item It is tempting to treat $\sqrt{\boldsymbol{y}+c}-\sqrt{\boldsymbol{\Phi x}+c}$ as a Gaussian random variable, and hence $R(\* y, \* \Phi \* x)$ as a chi random variable. This would ignore the fact that the Gaussianity of the former has been established only asymptotically if all the values in $\* \Phi \* x$ tend to $\infty$ \cite{Curtiss1943}. However we have in practice seen that even for moderate values of $\* \Phi \* x$, its distribution can be approximated very closely by a Gaussian as affirmed by Kolmogorov Smirnov hypothesis tests \cite{suppcode}, even though we are unable to prove this theoretically. In fact, we have found no literature that establishes even the sub-Gaussianity or sub-exponentiality of $R(\* y, \* \Phi \* x)$. Nonetheless, treating this approximation as exact allows us to improve the probability in the second part of the theorem from $1-1/N$ (for $\kappa = 1$) to $1-2e^{-N \tau}$ for an appropriately defined constant $\tau$. If we treat $\varepsilon$ as equal to the magnitude of a vector with elements drawn from $\mathcal{N}(0,\frac{1}{4})$, then $\varepsilon^2$ follows a chi distribution with $N$ degrees of freedom. Hence, we can use tail bounds on the chi-square random variable \cite{Laurent2000} (Lemma 1) to arrive at the following bound:
\begin{eqnarray} 
P \Big( \frac{\|\boldsymbol{\theta}-\boldsymbol{\theta^\star}\|_2}{I} \leq C_1\sqrt[]{N\tilde{\tau}} \sqrt{\frac{1}{I}+\frac{cN}{I^2}} \nonumber \\
+ \frac{C_2s^{-\frac{1}{2}}\|\boldsymbol{\theta}-\boldsymbol{\theta_s}\|_1}{I} \Big) \geq 1-\exp(-N\tau) \nonumber
\end{eqnarray}
for some $\tau > 0$ where $\tilde{\tau} \triangleq (1+2\tau+\sqrt{2\tau})$.
\end{enumerate}

\subsection{Theorem for Residual Magnitude in the Poisson-Gaussian case}
\begin{figure*}[!t]
\centering
\includegraphics[width=1.75in]{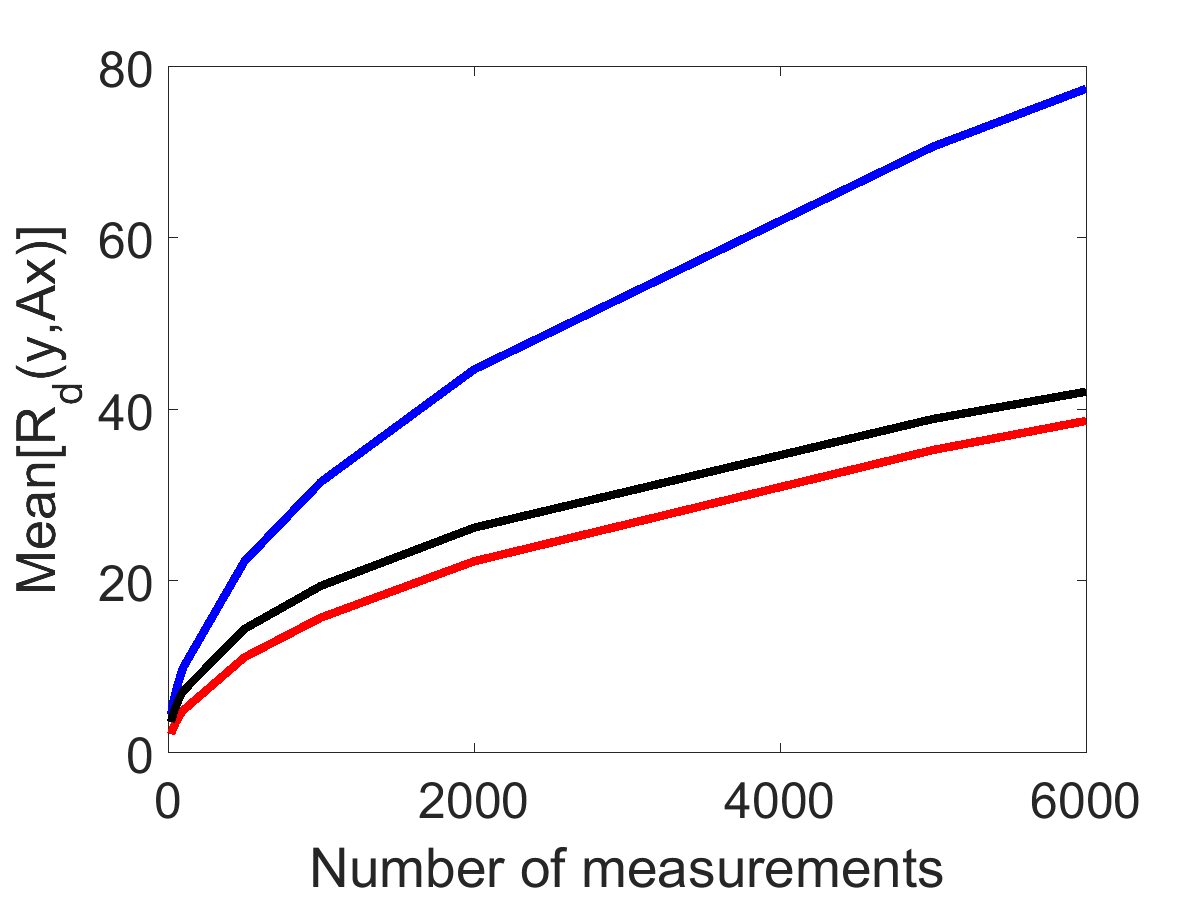}
\includegraphics[width=1.75in]{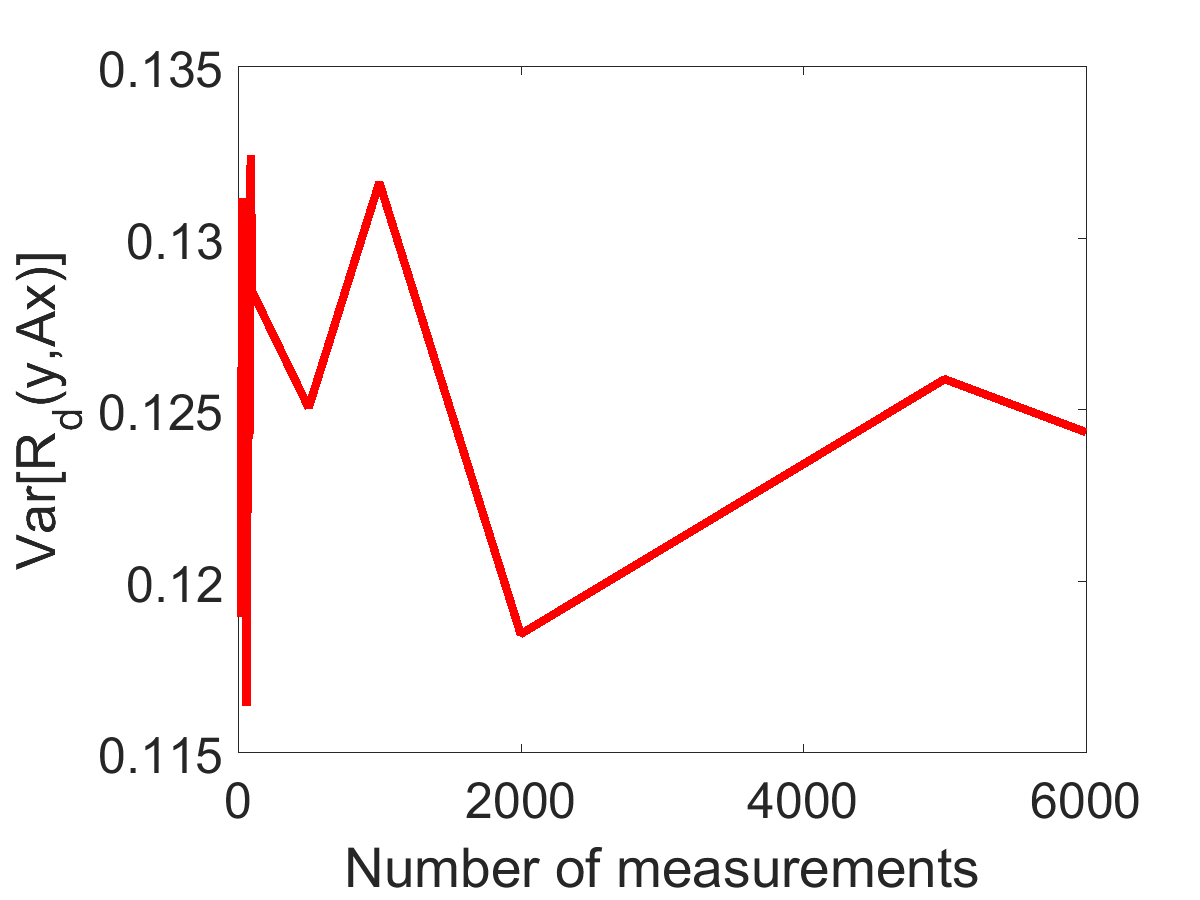}
\includegraphics[width=1.75in]{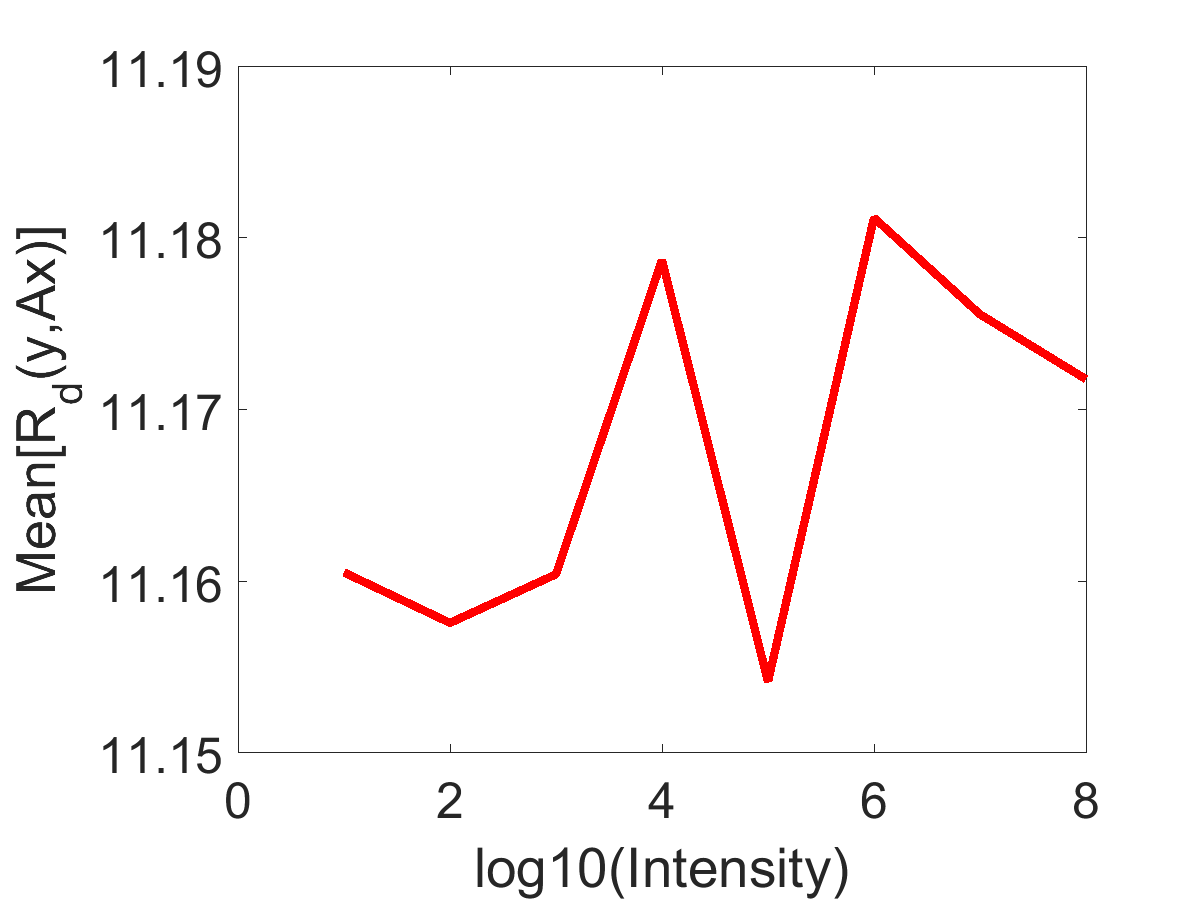}
\includegraphics[width=1.75in]{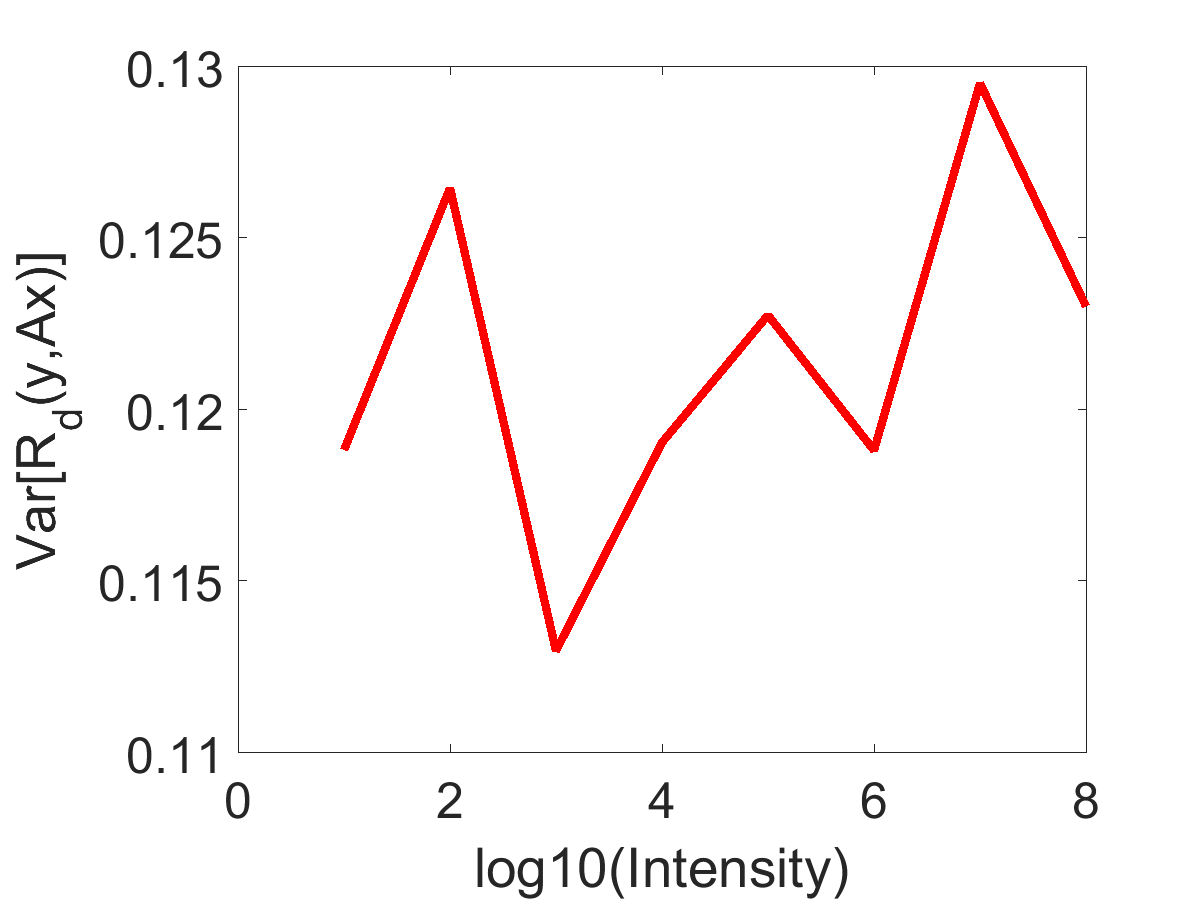}
\includegraphics[width=1.75in]{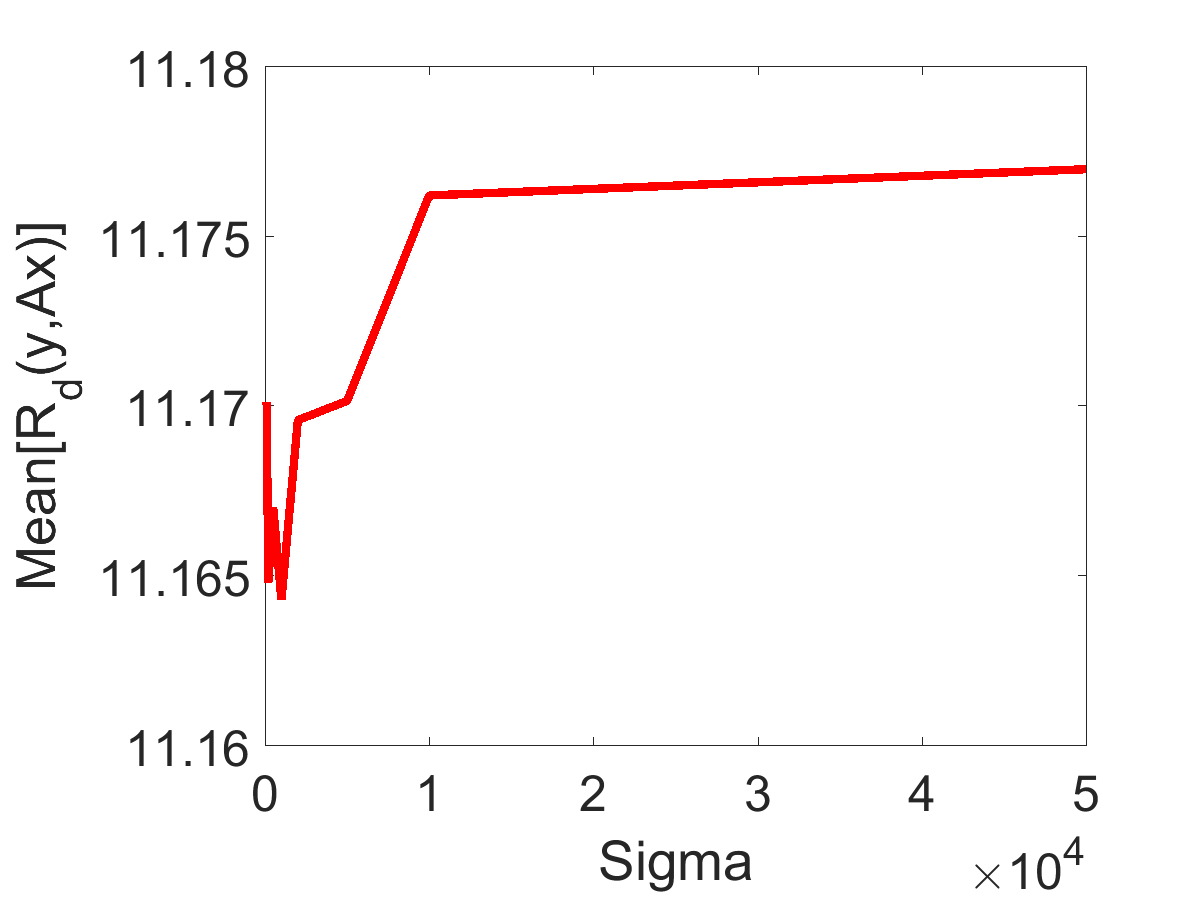}
\includegraphics[width=1.75in]{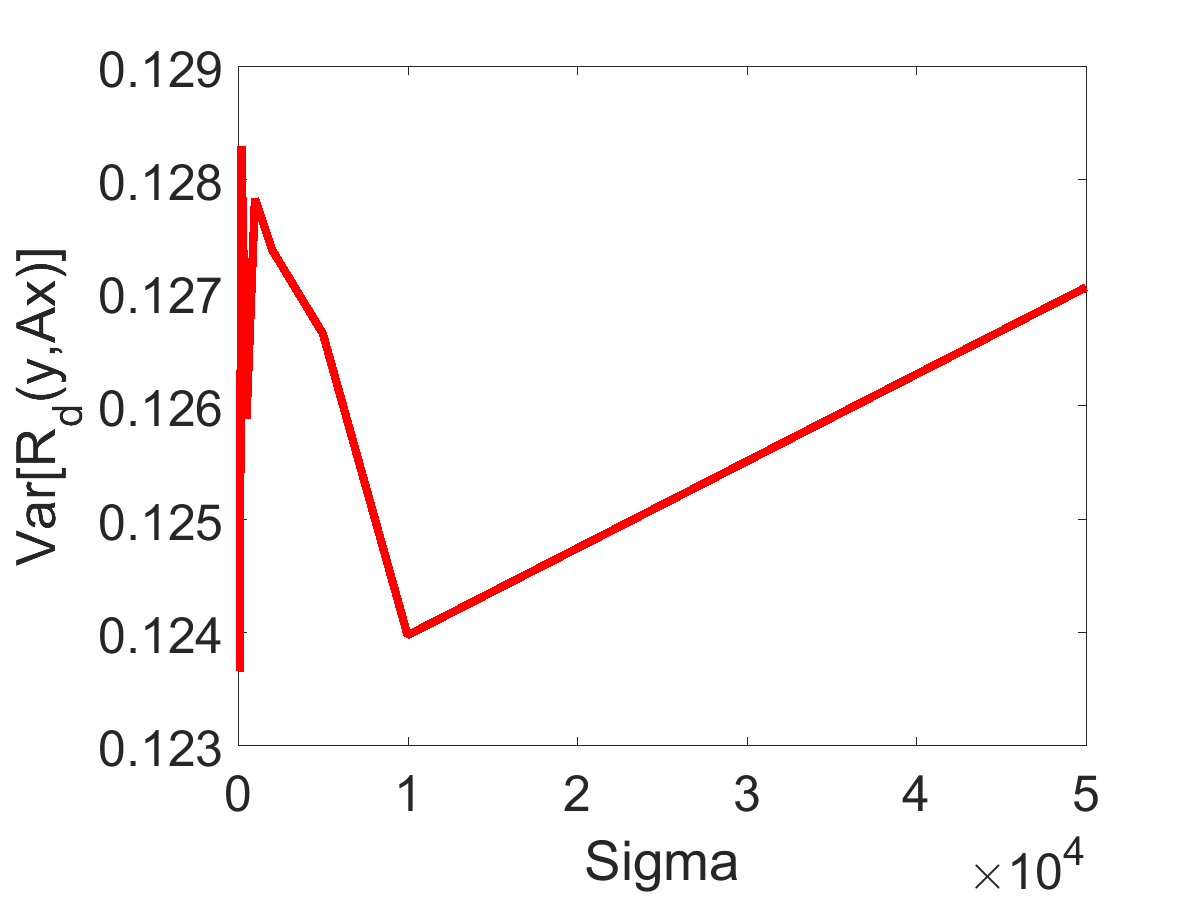}
\includegraphics[width=1.75in]{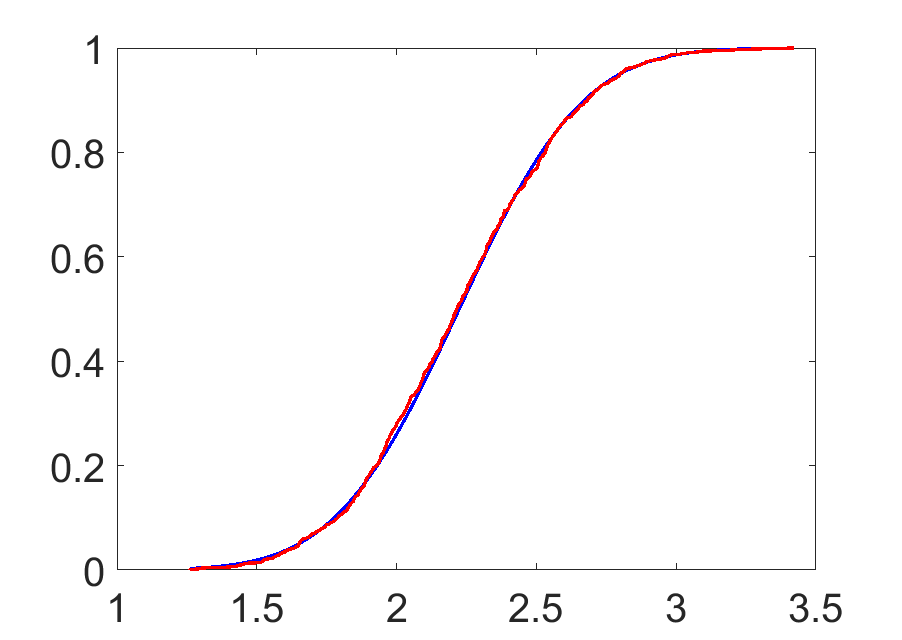} 
\caption{In the left to right, top to bottom order. First two sub-figures: Plot of mean and variance of the values of $R_d(\boldsymbol{y},\boldsymbol{\Phi x})$ versus $N$ for a fixed $I = 10^3, \sigma = 200$ for a signal of dimension $m = 1000$. (For the left subfigure in the first row, the blue line represents the plot of $\sqrt{N}$ and the black line represents $N^{0.43}$.) Third and fourth sub-figures: Plot of mean and variance of the values of $R_d(\boldsymbol{y},\boldsymbol{\Phi x})$ versus $\log_{10}(I)$ for a fixed $N=500, \sigma = 200$ for a signal of dimension $m = 1000$. Fifth and sixth sub-figures: Plot of mean and variance of the values of $R_d(\boldsymbol{y},\boldsymbol{\Phi x})$ versus $\sigma$ for a fixed $I = 10^3, N = 50$ for a signal of dimension $m = 1000$. Last sub-figure: Empirical CDF of $R_d(\boldsymbol{y},\boldsymbol{\Phi x})$ (red curve) for $N = 20, I = 10^3, m = 1000$ compared to a Gaussian CDF (blue curve) with mean and variance equal to that of the values of $R_d(\boldsymbol{y},\boldsymbol{\Phi x})$. The curves overlap significantly as the empirical CDFs are very close. Scripts for reproducing these results are available at \cite{suppcode}.}
\label{fig:Ganscombe_mean_var}
\end{figure*}

Here, we state a theorem for the case of Poisson-Gaussian noise in the compressed measurements (with a known standard deviation for the Gaussian part of the noise), equivalent to Theorem 1 for Poisson noise. The proof can be found in Section \ref{subsec:thm3}. This theorem is inspired by experimentally observed behaviour of $R_d(\boldsymbol{y},\boldsymbol{\Phi x}) \triangleq \|\sqrt{\boldsymbol{y}+d}-\sqrt{\boldsymbol{\Phi x}+d}\|_2$ where $d \triangleq c + \sigma^2$, which was quite similar to the Poisson case. That, is the mean of $R_d(\boldsymbol{y},\boldsymbol{\Phi x})$ appeared to be $\mathcal{O}(\sqrt{N}$ and the variance was a constant independent of $N, I, \sigma$. This can be seen in Fig. \ref{fig:Ganscombe_mean_var}.

\textbf{Theorem 3 :} Let $\boldsymbol{y}$ be a vector of $N$ independent CS measurements such that $y_i \sim \textrm{Poisson}[(\boldsymbol{\Phi x})_i] + \eta_i$ where $\boldsymbol{\Phi} \in \mathbb{R}^{N \times m}$ is a non-negative flux-preserving matrix as per Eqn. \ref{eq:Phi}, $\boldsymbol{x} \in \mathbb{R}^m$ is a non-negative signal and $\eta_i \sim \mathcal{N}(0,\sigma^2)$. Define $\gamma_i \triangleq (\boldsymbol{\Phi x})_i$, $d \triangleq c + \sigma^2$ and $R_d(\boldsymbol{y},\boldsymbol{\Phi x}) \triangleq \|\sqrt{\boldsymbol{y}+d}-\sqrt{\boldsymbol{\Phi x}+d}\|_2$. Then we have:
\begin{enumerate}
\item $E[R_d(\boldsymbol{y},\boldsymbol{\Phi x})] \leq \sqrt{N/2}$
\item Define $v \triangleq \textrm{Var}[R_d(\boldsymbol{y},\boldsymbol{\Phi x})]$. Then we have \\
$v \leq \dfrac{\sum_{i=1}^N \frac{\gamma_i(1+3\gamma_i)+\sigma^4}{(\gamma_i+d)^2}} {\sum_{i=1}^N \textrm{max}(0,\frac{\gamma_i}{4(\gamma_i+d)}-\frac{\gamma_i}{8(\gamma_i+d)^2})}$
\item If $\forall i, \gamma_i \geq 1$, we see that $v \leq v_u \triangleq \frac{1.25}{(2d+1)/(8(1+d)^2)} $
\item $P\Big(R_d(\* y, \*\Phi\*x) \leq \sqrt{N}(\frac{1}{\sqrt{2}}+\sqrt{v_u})\Big) \geq 1-1/N$.
\end{enumerate}
We make a few comments below:
\begin{enumerate}
\item Yet again, $E[R_d(\boldsymbol{y},\boldsymbol{\Phi x})]$ does not increase with $I$. This property is \emph{not} shared by the negative log-likelihood of the Poisson-Gaussian distribution. Also when $\boldsymbol{\Phi x} \succeq \boldsymbol{1}$, we see that $\textrm{Var}[R_d(\boldsymbol{y},\boldsymbol{\Phi x})]$ is again a constant dependent only on $\sigma$.
\item In practice, we observed that $\textrm{Var}[R_d(\boldsymbol{y},\boldsymbol{\Phi x})]$ did not depend even on $\sigma$ (see Fig. \ref{fig:Ganscombe_mean_var}). This particular bound is slightly loose because of inequalities used in various steps of our proof which gave rise to an extra $\sigma$ term. 
\item Setting $\sigma = 0$ produces the statement of Theorem 1.
\item The bounds in this theorem can be easily modified for the case of uniform quantization noise from $\textrm{Unif}[-\delta,+\delta]$, or Gaussian noise coupled with uniform quantization noise. 
\end{enumerate}

\subsection{Key Theorem for Poisson-Gaussian CS}
For the Poisson-Gaussian case, a theorem similar to Theorem 2 follows. \\
\textbf{Theorem 4 :} Consider a non-negative signal $\boldsymbol{x}$ with total intensity $I \triangleq \|\boldsymbol{x}\|_1$ expressed using the orthornormal basis $\boldsymbol{\Psi}$ in the form $\boldsymbol{x} = \boldsymbol{\Psi\theta}$. Consider Poisson-Gaussian corrupted CS measurements of the form $\boldsymbol{y} \sim \textrm{Poisson}(\boldsymbol{\Phi x}) + \boldsymbol{\eta}$ where $\boldsymbol{\eta} \sim \mathcal{N}(0,\sigma^2)$ is signal-independent noise, and $\boldsymbol{\Phi}$ is constructed as per Eqn. \ref{eq:Phi}. Let $\boldsymbol{\theta^\star}$ be the result of the following optimization problem:
\begin{eqnarray}
(PG2): \textrm{min} \|\boldsymbol{\theta}\|_1 \textrm{ such that } \|\sqrt{\boldsymbol{y}+d}-\sqrt{\boldsymbol{A \theta}+d}\|_2 \leq \varepsilon, \\ \nonumber
\|\boldsymbol{\Psi \theta}\|_1 = I, \boldsymbol{\Psi \theta} \succeq \boldsymbol{0},
\end{eqnarray}
where $d \triangleq c + \sigma^2$, $\*A \triangleq \*\Phi \*\Psi$ so that $\*\Phi \*x = \*A\*\theta$, $\varepsilon \triangleq \sqrt{N}(\sqrt{v_u} + \frac{1}{\sqrt{2}})$ is an upper bound on the magnitude of the noise in the measurements \emph{after} application of the GAT. Let $\boldsymbol{\theta_s}$ denote a vector containing the $s$ largest magnitude elements of $\boldsymbol{\theta}$ with the rest being 0. If $\boldsymbol{\widetilde \Phi}$ obeys RIP of order $2s$ with RIC $\delta_{2s} < \sqrt{2}-1$, and $\boldsymbol{\Phi x} \succeq \* 1$, then we have for any $\kappa > 0$:
\begin{align*}
%E \bigg( \frac{\|\boldsymbol{\theta}-\boldsymbol{\theta^\star}\|_2}{I} \bigg) \leq C_1\sqrt{\dfrac{N}{2}} \sqrt{\frac{1}{I}+\frac{dN}{I^2}} + \frac{C_2s^{-\frac{1}{2}}\|\boldsymbol{\theta}-\boldsymbol{\theta_s}\|_1}{I},\\
P \bigg( \frac{\|\boldsymbol{\theta}-\boldsymbol{\theta^\star}\|_2}{I} \leq C_1\sqrt{N}\tau_d \sqrt{\frac{1}{I}+\frac{dN}{I^2}} + \frac{C_2s^{-\frac{1}{2}}\|\boldsymbol{\theta}-\boldsymbol{\theta_s}\|_1}{I} \bigg) \\ \nonumber
\geq 1-\kappa^2/N \textrm{ where } \tau_d \triangleq (\sqrt{v_u}/\kappa+\frac{1}{\sqrt{2}}).
\end{align*}
\\
\textbf{Remarks on Theorem and its Proof:}
\begin{enumerate}
\item The proof of this theorem follows Theorem 2 very closely with a replacement of $c$ by $d$. Hence we omit its proof.
\item Theorem 2 and Theorem 4 show that using the VST, a unified treatment of Poisson CS as well as Poisson-Gaussian CS is possible. Methods based on purely the negative Poisson log-likelihood do not have this feature. Theorem 4 can be easily extended to include uniform quantization noise (with or without Gaussian noise). 
\item For the same probability, the upper bounds increase with $\sigma$ due to the $d$ term in the square root. Also setting $\sigma = 0$ gives us Theorem 2.
\item Similar to the case of Theorem 2, the constant factors in the bounds can be approximately refined using the CLT for large $N$. 
\end{enumerate}

\subsection{Properties of $R(\*y,\*\Phi \*x)$ and $R_d(\*y,\*\Phi \*x)$}
First, we note $R^2(\*y,\*\Phi \*x)$ is convex in $\*x$, which can be seen by a simple algebraic expansion and due to the concavity of $\sqrt{\*x}$. Also, it is convex in $\*\theta$ due to the affine mapping property of convex functions (see Section 3.2.2. of \cite{Boyd_book}). Second, for finite $\*y$ and $c \neq 0$, $R^2(\*y,\*\Phi \*x)$ is Lipschitz continuous as it has a bounded first derivative. Both these properties are also true for $R^2_d(\*y,\*\Phi \*x)$. These properties allow for efficient optimization and have been pointed out earlier in \cite{Dupe2009}.

\section{Results}
\label{sec:results}
In this section, we show signal reconstruction results from CS measurements with Poisson and Poisson-Gaussian noise. Box-plots for the results of all these experiments are presented in the supplemental material accompanying this paper. Our scripts for reproducing the results in this section are available at \cite{suppcode}.

\subsection{Experiments on Poisson CS}
\label{subsec:results_P}
\textit{Signal and Measurement Generation:} We ran reconstruction experiments on reconstruction of $Q = 100$ non-negative signals in 1D with 100 elements each, from their Poisson corrupted CS measurements. The sensing matrix $\*\Phi$ followed Eqn. \ref{eq:Phi}. The signals were synthetically constructed using sparse linear combinations of DCT basis vectors.  The non-zero indices of the coefficient vector $\boldsymbol{\theta}$ for the $Q$ different signals were chosen randomly (i.e. allowing different supports for each signal), and the values of those entries were drawn randomly from $\textrm{Unif}[0,1]$. The signals $\*x =\*\Psi \*\theta$ thus generated were forced to be non-negative by adjusting the DC component, followed by a scaling to ensure that they had a desired value of $I$ (see description of experiments later in this section). \\
\textit{Methods Compared:} For the Poisson noise case, we ran our simulations on the following problem which is a variant of (P1) \emph{without} the constraint $\|\boldsymbol{\Psi \theta}\|_1 = I$ as its exclusion had a negligible impact on the results (see later in this section): 
\begin{equation}
(P3): \textrm{min} \|\boldsymbol{\theta}\|_1 \textrm{ such that } \|\sqrt{\boldsymbol{y}+c}-\sqrt{\boldsymbol{A \theta}+c}\|_2 \leq \varepsilon, \\ \nonumber
\boldsymbol{\Psi \theta} \succeq \boldsymbol{0}.
\end{equation}
Here we set $c = 3/8$, and the bound $\varepsilon$ was set to $2\sqrt{N}$ based on the tail bound from Theorem 1 (note that $2\sqrt{N} = \sqrt{N}/\sqrt{2} + \sqrt{N}(3.29/2.5)$, and that this bound holds with probability $1-(2.5)^2/N$, i.e. $\kappa=2.5$). Note that the same value of $\varepsilon$ was used in all experiments, and that this is a \emph{very conservative} upper bound. Problem (P3), being convex, was implemented using the well-known CVX package \cite{CVX2014} with the SDPT3 solver. We compared the performance of (P3) to the following problem based on the negative log-likelihood of the Poisson distribution (again without the constraint $\|\boldsymbol{\Psi \theta}\|_1 = I$ for the same reason as for (P3)):
\begin{equation}
(P4): \textrm{min}\hspace{0.1cm} \rho \|\boldsymbol{\theta}\|_1 + \sum_{i=1}^N ((\boldsymbol{A \theta})_i - y_i \log (\boldsymbol{A \theta})_i), \\ \nonumber
\boldsymbol{\Psi \theta} \succeq \boldsymbol{0}.
\end{equation}
For (P4), the regularization parameter $\rho$ was chosen omnisciently from the set $\mathcal{S} \triangleq \{10^{-10},10^{-9},...,10\}$, i.e. choosing the particular value of $\rho \in \mathcal{S}$ that yielded the least squared difference between the true $\boldsymbol{\theta}$ (assuming it were known) and its estimate. (P4) was implemented using the well-known SPIRAL-TAP algorithm \cite{Harmany2012} with a penalty for the $\ell_1$ norm of DCT coefficients, for a maximum of 500 iterations (in many cases, the algorithm converged and exited in just 300-400 iterations). For the default choice of a maximum of 100 iterations set in the SPIRAL-TAP code, the RRMSE increased significantly. We used default choices for all other parameters except $\rho$. Additionally, we also compared the results to a version of (P3) which we had used in \cite{Garg2017}, given by the following:
\begin{equation}
(P5): \textrm{min}\hspace{0.1cm} \rho \|\boldsymbol{\theta}\|_1 + \|\sqrt{\boldsymbol{y}+c}-\sqrt{\boldsymbol{A \theta}+c}\|^2_2, \\ \nonumber
\boldsymbol{\Psi \theta} \succeq \boldsymbol{0},
\end{equation}
where $\rho$ was chosen omnisciently from $\mathcal{S}$. (P5), being convex, was again implemented using CVX and SDPT3. 
\begin{figure}
	\includegraphics[width=0.8\linewidth]{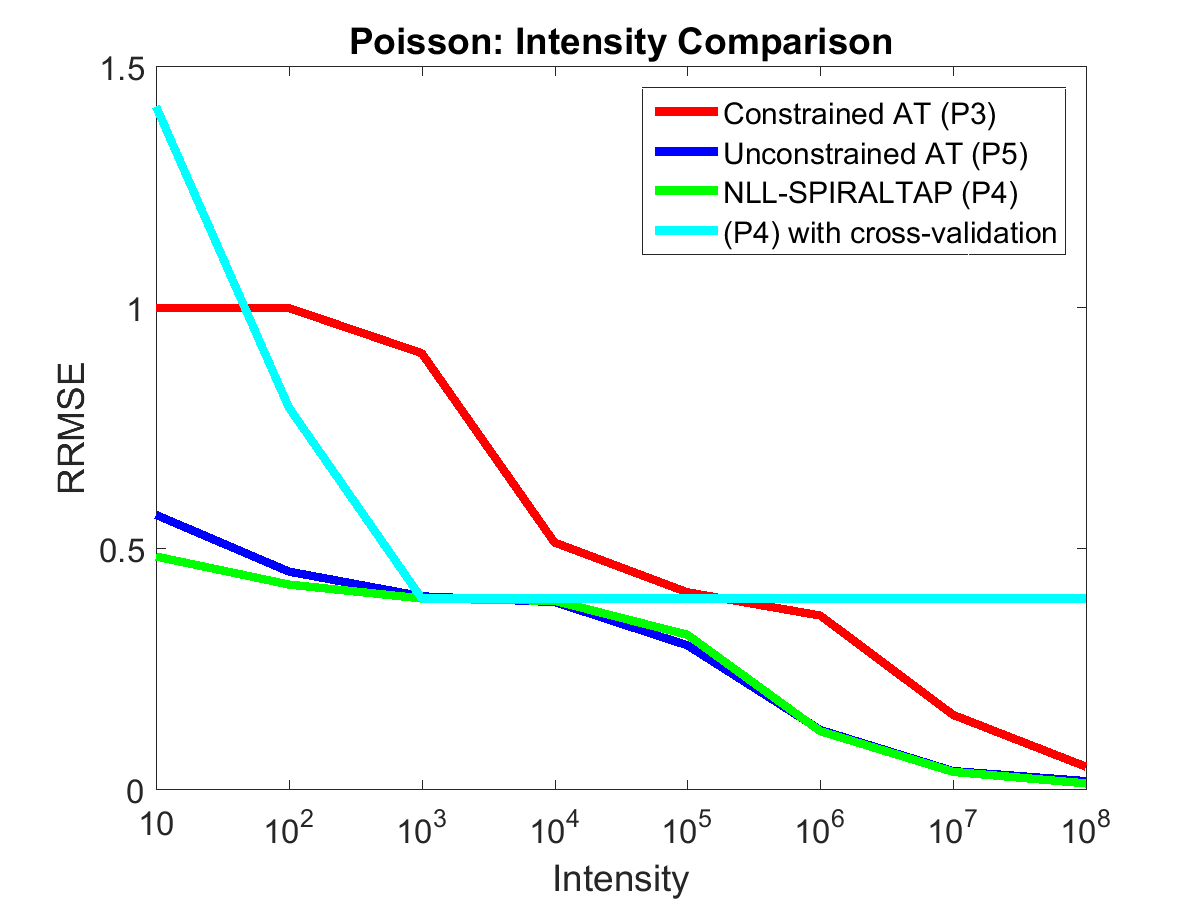}	
    \includegraphics[width=0.8\linewidth]{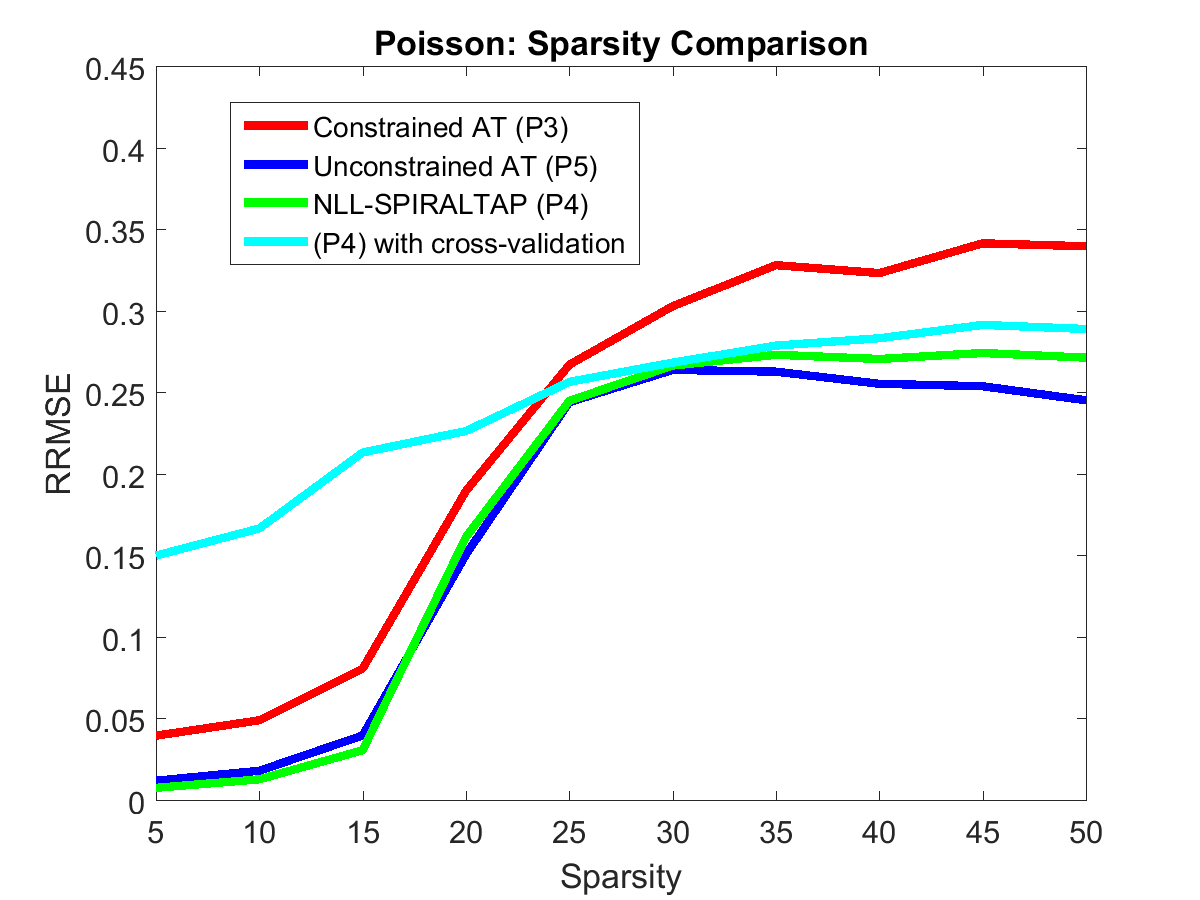}
	\includegraphics[width=0.8\linewidth]{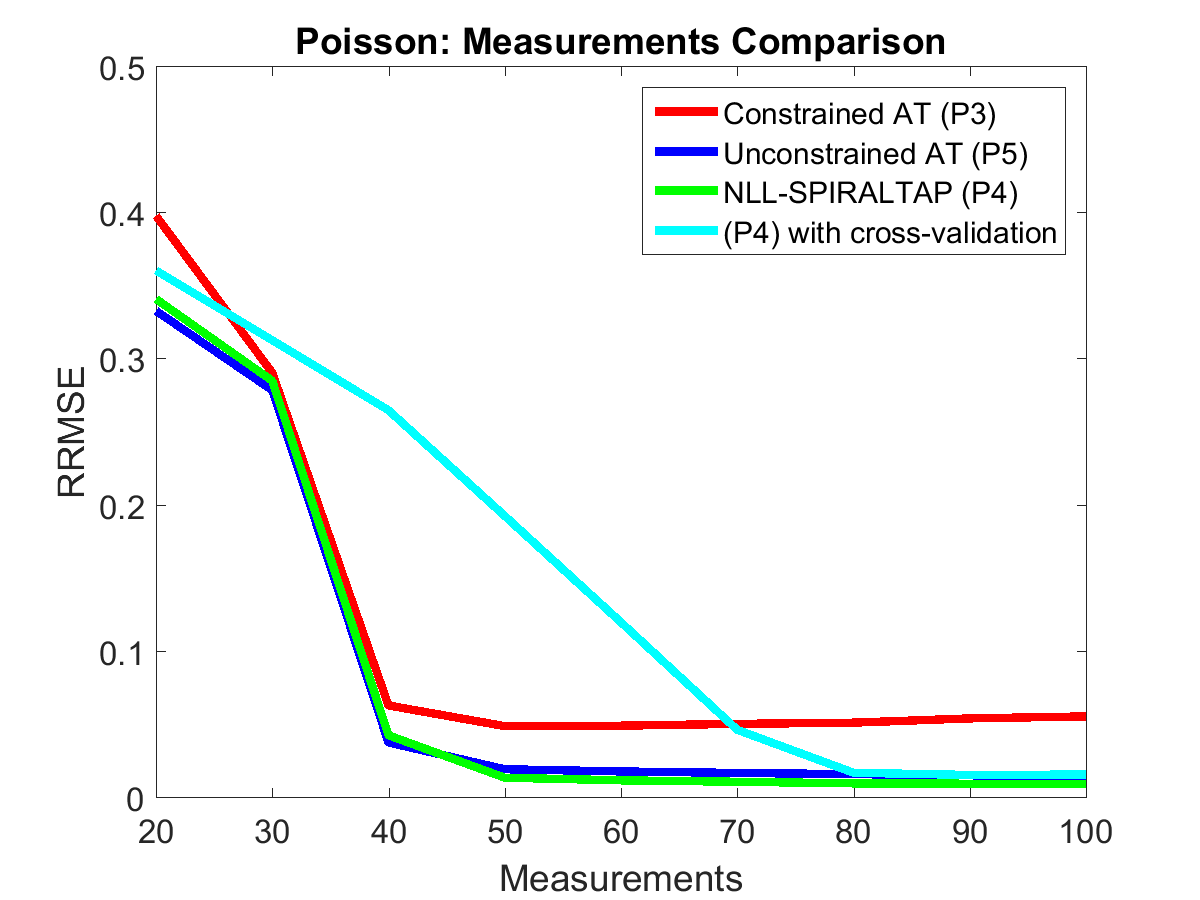}    
  	\caption{Median RRMSE comparisons between (P3) using CVX with $\varepsilon = 2\sqrt{N}$ (termed `Constrained Anscombe'), (P4) using SPIRAL-TAP (termed `NLL SPIRAL-TAP) with omniscient $\rho$, (P4) with cross-validation for $\rho$, and (P5) using CVX (termed `Unconstrained Anscombe'). Top row: fixed $N=50$ and $s=10$ but varying $I$, middle row: fixed $I=10^8$ and $N=50$ but varying $s$, bottom row: fixed $I=10^8$ and $s=10$ but varying $N$. See supplemental material for box-plots and \cite{suppcode} for code.}
    \label{fig:comparisons}
\end{figure}

\textit{Study of variation of signal/measurement parameters:} We show comparisons between (P3), (P4) with SPIRAL-TAP, and (P5) for three types of experiments for the following RRMSE (relative root mean-squared error) metric: $\textrm{RRMSE} = \|\*x-\*x^{\star}\|_2/\|\*x\|_2$, where $\*x$ and $\*x^{\star}$ denote the true/original and reconstructed signal respectively. In the \textit{first experiment}, we studied the effect of change in signal intensity $I$ on the reconstruction results. For this, we generated Poisson corrupted measurements of the $Q$ different signals in $\mathbb{R}^{100}$, each with a fixed number of measurements $N = 50$. The sparsity of each signal in the DCT basis was fixed to $s = 10$ (but with different supports), and the signal intensity was varied from $I=10$ to $I=10^{8}$ in powers of 10. For each value of $I$, the median RRMSE value over the $Q$ signals was computed. This is shown in the top sub-figure in Fig. \ref{fig:comparisons}. The performance of all methods improves with increase in $I$ as expected. In the \textit{second experiment}, for the $Q$ different signals, the number of Poisson corrupted CS measurements was fixed to $N = 50$, the signal intensity was fixed to $I = 10^8$, and the signal sparsity was varied from $s = 5$ to $s = 50$ in steps of 5. For each value of $s$, median RRMSE values were recorded over the $Q$ signals, as shown in the middle sub-figure in Fig. \ref{fig:comparisons}. The performance of all methods worsens with increase in $s$ as expected. In the \textit{third experiment}, for the $Q$ different signals, the sparsity of the signals was fixed to $s = 10$, and their intensity was fixed to $I = 10^8$. The number of measurements was varied from $N = 20$ to $N = 100$ in steps of 10. For each value of $N$, median RRMSE values were recorded over the $Q$ signals, as shown in the bottom sub-figure in Fig. \ref{fig:comparisons}. We do see an improvement in the reconstruction results with increase in $N$, but this is not guaranteed in the worst case similar to \cite{Raginsky2010}. 

\textit{Observations and Comments:} Observing Fig. \ref{fig:comparisons}, we see that the reconstruction results with (P5) and (P4) are comparable in most cases. (P5) and (P4) showed better results than (P3) due to the omnisicent selection of $\rho$, as against the fixed, statistically motivated $\varepsilon$ in (P3). Note that omniscient choices are difficult to implement in practice, and have significant computational costs. Improper choice of $\rho$ led to arbitrary increase in reconstruction error. We have found that the optimal $\rho$ depended on the unknown signal (see also \cite{YHLi2015} and Table \ref{tab:comparisons}). While model-selection approaches for Poisson problems exist \cite{Bardsley2009}, no performance bounds with such methods have been proven. For the sake of comparison, we collected results on (P4) via cross-validation. For this, we omnisciently chose $\rho$ which yielded the best RRMSE for $I = 10^4$ and used the same $\rho$ for all other intensity levels in the first experiment. For the second experiment, $\rho$ was chosen omnisciently for $s = 30$ and used for all other values of $s$. For the third experiment, $\rho$ was chosen omnisciently for $N = 20$ and used for all other $N$. The results for this variant of (P4) (termed `(P4) with cross-validation') are shown in Fig. \ref{fig:comparisons}. Recent work in \cite{Rohban2016} analyzed the following estimator instead of (P4) for $\boldsymbol{\Psi} = \boldsymbol{I_d}$ (identity basis):
\begin{equation}
(P6): \textrm{min}_{\|\boldsymbol{\theta}\|_1 \leq I, \boldsymbol{\theta} \succeq \boldsymbol{0}} \sum_{i=1}^N (\boldsymbol{A \theta})_i - y_i \log (\boldsymbol{A \theta})_i.
\end{equation}
The method requires prior knowledge of $I$ for the analysis as well as the implementation even for matrices that obey RIP. In our case, as also in \cite{Jiang2015,Raginsky2010,Jiang2015_arxiv}, the constraint $\|\boldsymbol{x}\|_1 = I$ is required in the theoretical analysis for the specific type of matrices from Eqn. \ref{eq:Phi}. The constraint would not be required for RIP-obeying matrices, and was not deemed necessary even in the numerical experiments for matrices from Eqn. \ref{eq:Phi}. For example, RRMSE of a typical signal of 100 dimensions with $s = 10, I = 10^8$ with $N = 50$ CS measurements using (P3) was greater than that using (P1) by only $\mathcal{O}(10^{-4})$. 

\textit{Execution Times:} We also saw that (P4) for a single fixed $\rho$ (that is, not counting execution times for different $\rho \in \mathcal{S}$) was 3-4 times more computationally expensive than (P3) with a fixed $\varepsilon$. On a 2GHz CPU with 8 GB RAM, typical execution times were 58 seconds and 18.6 seconds for (P4) and (P3) respectively, for $N = 50, m = 100, s = 10$. 

\begin{figure}
	\includegraphics[width=0.3\linewidth]{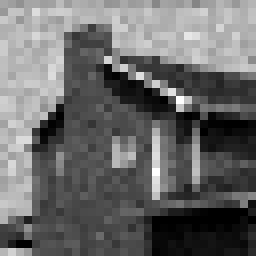}	
    \includegraphics[width=0.3\linewidth]{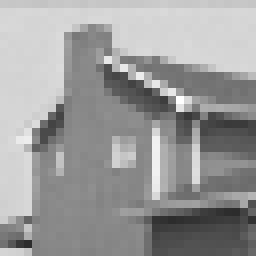}
	\includegraphics[width=0.3\linewidth]{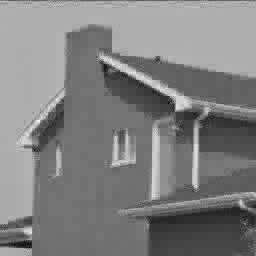}    

	\includegraphics[width=0.3\linewidth]{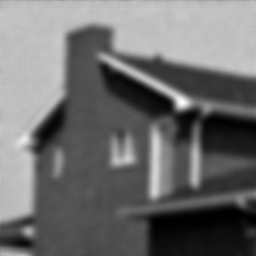}	
    \includegraphics[width=0.3\linewidth]{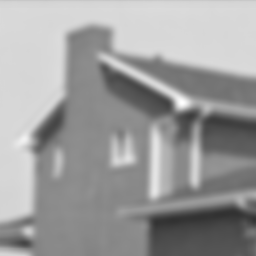}
	\includegraphics[width=0.3\linewidth]{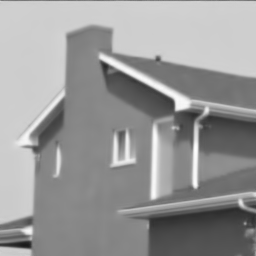}    

	\includegraphics[width=0.3\linewidth]{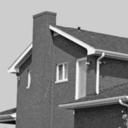}    
  	\caption{First row: Image reconstruction results for non-overlapping $8 \times 8$ patches from 32 CS measurements per patch, using (P3) for $I = 10^6$ (left, RRMSE = $0.743$), $I = 10^8$ (middle, RRMSE = $0.16$) and $I = 10^{10}$ (right, RRMSE = $0.068$). Second row: Same as in the first row but with overlapping patches and averaging in sliding window fashion: for $I = 10^6$ (left, RRMSE = $0.7408$), $I = 10^8$ (middle, RRMSE = $0.148$) and $I = 10^{10}$ (right, RRMSE = $0.054$). Third row: original image for reference.}
    \label{fig:imagerecon_Poiss}
\end{figure}

\textit{Image Reconstruction:} Lastly, we ran an experiment to simulate image-patch and image reconstruction from Poisson-corrupted CS measurements, for a camera following the architecture of \cite{Oike2013},\cite{Kulkarni2016}. The architecture of these cameras is similar to the Rice SPC \cite{Duarte2008}, but the measurements are acquired patch-wise. That is, for each patch $\*x_i \in \mathbb{R}_{+}^m$ extracted from an image, the measurement vector is given by $\*y_i \sim \textrm{Poisson}(\*\Phi_i \*x_i)$ where $\*y_i \in \mathbb{Z}_{+}^N, \* \Phi_i \in \mathbb{R}_{+}^{N \times m}, N \ll m$ and $i$ is a spatial location index. The model for each $\* \Phi_i$ follows Eqn. \ref{eq:Phi}. In our experiments, we set $m = 64$ (from $8 \times 8$ patches) and $N = 32$. Each (non-overlapping) patch $\*x_i$ was independently reconstructed by solving (P3) using $\*\Psi$ as the 2D-DCT basis and $\varepsilon = 2\sqrt{N}$, as per the tail bound on $R(\*y,\*\Phi \*x)$. Since there are inevitable patch-seam artifacts, we also ran these experiments for overlapping patches followed by sliding-window averaging. Though in \cite{Oike2013},\cite{Kulkarni2016}, CS measurements are not acquired on overlapping blocks, this simulates the use of a deblocking algorithm to get rid of patch-seam artifacts. The reconstruction results for this experiment are presented in Fig. \ref{fig:imagerecon_Poiss} on the popular `house' image (size $256 \times 256$) for values of total image-intensity $I \in \{10^6, 10^8, 10^{10}\}$. The results show clear improvement with increase in $I$ and are evidence that our method works for compressible signals as well, since image patches are compressible (not sparse) in 2D-DCT bases.

\subsection{Experiments on Poisson-Gaussian CS}
The signal generation model for experiments on Poisson-Gaussian CS was the same as that used for Poisson CS. Throughout, we assumed known values of $\sigma$. Experiments were performed for the problem (PG3) defined below, which is identical to (PG2) except that we did not impose the $\|\boldsymbol{x}\|_1 = I$ constraint as its exclusion had negligible impact on the results:
\begin{eqnarray}
(PG3): \textrm{min} \|\boldsymbol{\theta}\|_1 \textrm{ s.t. } \|\sqrt{\boldsymbol{y}+d}-\sqrt{\boldsymbol{A \theta}+d}\|_2 \leq \varepsilon, \boldsymbol{\Psi \theta} \succeq \boldsymbol{0}.
\end{eqnarray}
Here as defined before $d \triangleq c + \sigma^2, c = 3/8$. For \emph{all} experiments using (PG3), the bound $\varepsilon$ was set to $2\sqrt{N}$ based on Fig. \ref{fig:Ganscombe_mean_var}. (The tail bound on $R_d(\*y,\*\Phi \*x)$ in Theorem 3 is loose by a factor of $\sigma$. Nevertheless, $2\sqrt{N}$ remains a very conservative upper bound.) We removed all measurements $y_i$ for which $y_i + d < 0$. This happened very rarely, and is akin to the so-called `saturation rejection'  for CS with saturation and quantization \cite{Laska2011}. (PG3) was implemented using CVX and the SDPT3 solver. We compared the results for (PG3) with those produced by problem (P4). (P4) was implemented using SPIRAL-TAP for a maximum of 500 iterations (ensuring convergence in each case) under default parameters except $\rho$ which was chosen omnisciently from $\mathcal{S}$. For (P4), all negative measurements were removed. We also compared the results with problem (PG5) defined below:
\begin{equation}
(PG5): \textrm{min}\hspace{0.1cm} \rho \|\boldsymbol{\theta}\|_1 + \|\sqrt{\boldsymbol{y}+d}-\sqrt{\boldsymbol{A \theta}+d}\|^2_2, \\ \nonumber
\boldsymbol{\Psi \theta} \succeq \boldsymbol{0}.
\end{equation}
(PG5) was implemented using CVX-SDPT3, using an omniscient choice of $\rho \in \mathcal{S}$ and with removal of measurements for which $y_i + d < 0$. We did not compare with the Poisson-Gaussian technique in \cite{Chouzenoux2015} because it is a \emph{deconvolution} algorithm with a total variation prior, whereas we are dealing with CS and sparsity of transform coefficients. We also observed that empirical results with AT (i.e. (P3)) were similar to those with GAT (i.e. (PG3)) for small to moderate values of $\sigma$. For larger $\sigma$, GAT outperformed AT, besides being statistically more principled. Moreover for AT, measurements for which $y_i + c < 0$ need to be removed. This occurs more often than $y_i + d < 0$ since $d \triangleq c + \sigma^2$. 
\begin{figure}
	\includegraphics[width=0.8\linewidth]{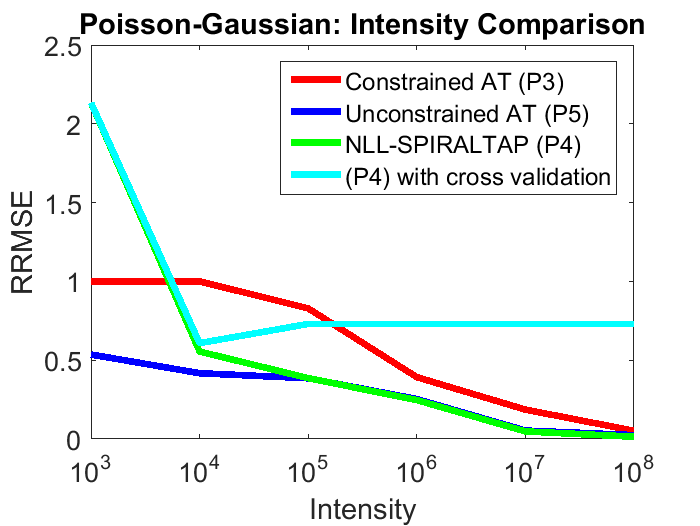}	
	\includegraphics[width=0.8\linewidth]{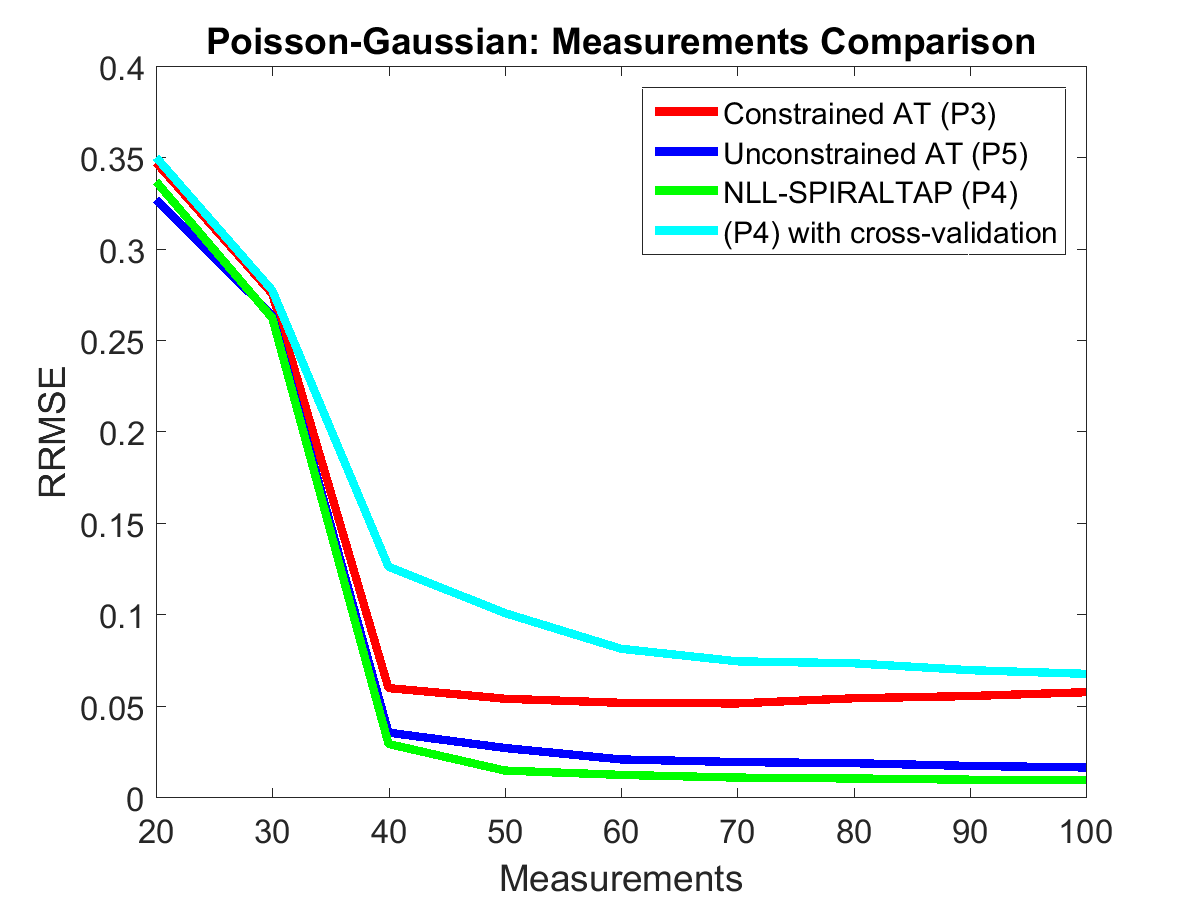}    
	\includegraphics[width=0.8\linewidth]{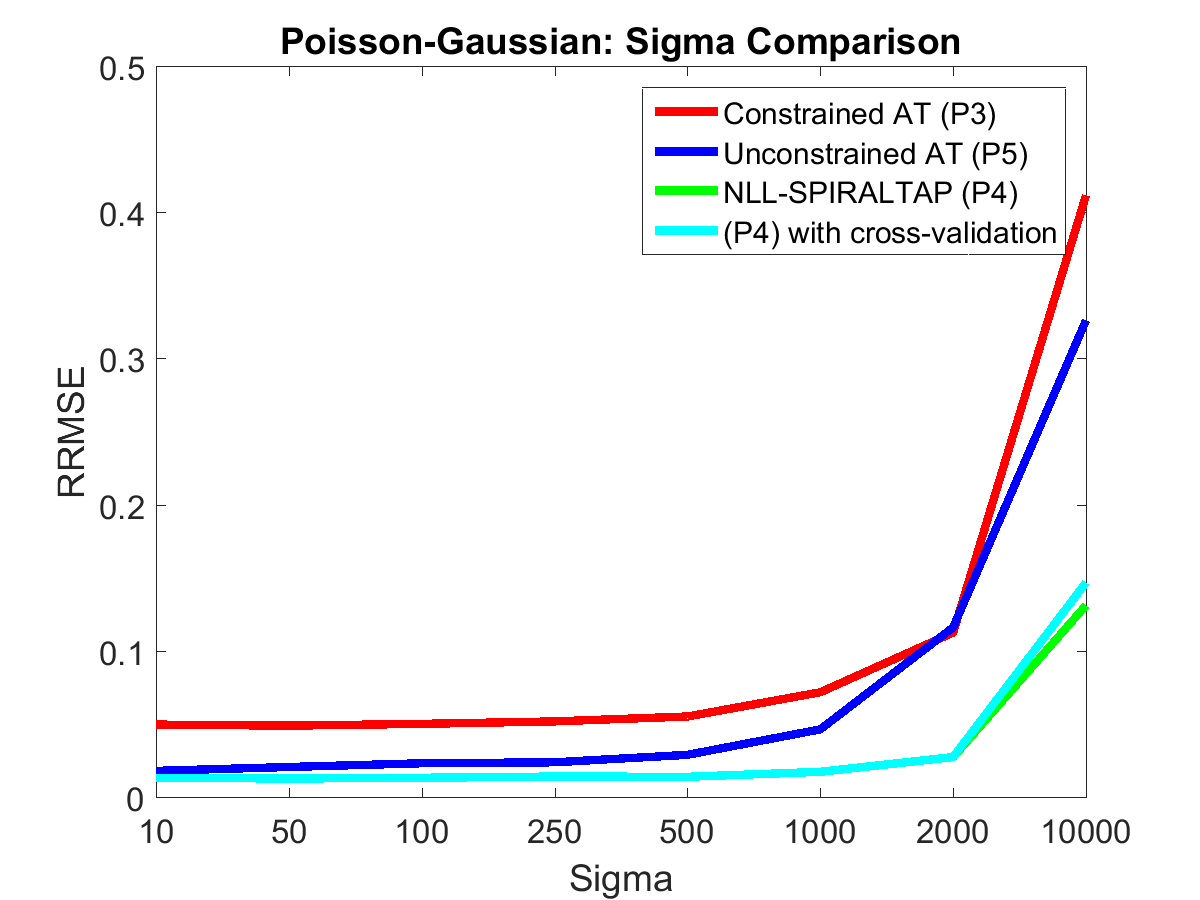}	
  	\caption{Median RRMSE comparisons between (PG3) using CVX with $\varepsilon = 2\sqrt{N}$ (termed `Constrained GAT'), (P4) using SPIRAL-TAP (termed `NLL SPIRAL-TAP'), (P4) with cross-validation for $\rho$, and (PG5) using CVX (termed `Unconstrained GAT'). Top row: fixed $N=50$, $\sigma=200$ and $s=10$ but varying $I$, middle row: fixed $I=10^8$, $s=10$ and $N=50$ but varying $\sigma$, bottom row: fixed $I=10^8$, $\sigma=200$ and $s=10$ but varying $N$. See supplemental material for box-plots and \cite{suppcode} for code.}
    \label{fig:PG_comparisons}
\end{figure}

\textit{Study of variation of signal/measurement parameters:} We ran three sets of experiments here. In the first experiment, we fixed $N = 50, s = 10, \sigma = 200$ and varied only $I$ from $10^3$ to $10^8$ in multiples of 10. In the second experiment, we fixed $I = 10^8, \sigma = 200, s = 10$ and varied only $N$ from $10$ to $100$ in steps of 10. In the third experiment, we fixed $I = 10^8, N = 50, s = 10$ and varied $\sigma$ in $\{10,50,100,250,500,1000,2000,10^4\}$. Comparative median RRMSE plots (across $Q$ signals) are presented in Fig. \ref{fig:PG_comparisons}. 

\textit{Observations and Comments:} The performance of our methods improved with increase in $I$ and $N$, and worsened gradually with increase in $\sigma$ (gradually because of the term $dN/I^2$ in the bounds for Theorem 4 which increases very slowly with $\sigma$ for large values of $I$, such as $I = 10^8$ as chosen in Fig. \ref{fig:PG_comparisons}). The presented results establish the usefulness of our proposed method for Poisson-Gaussian CS. We observed that (P4) and (PG5) with omnisicent $\rho$ outperformed (PG3) with fixed $\varepsilon$. Quite surprisingly, (P4) with omniscient $\rho$ performed very well, even though it is not designed for Poisson-Gaussian noise. However we emphasize that no theoretical performance bounds for (P4) have been established for this noise model. Moreover, with improperly chosen $\rho$, the performance of (PG5) and (P4) was worse than (PG3), and even for a single fixed $\rho$, (P4) was computationally more expensive than (P3). In Fig. \ref{fig:PG_comparisons}, we also show results for (P4) with cross-validation. In the first experiment, the value $\rho$ was omnisciently chosen for $I = 10^4$ and used for other intensities. In the second experiment, the value $\rho$ was chosen omnisciently for $N=30$ and used for other values of $N$. For the third experiment, we chose the best $\rho$ omnisciently for $\sigma = 20$ and used it for other values of $\sigma$. Surprisingly, the best $\rho$ did not depend on $\sigma$ for a wide range. 

\begin{figure}
    \includegraphics[width=0.3\linewidth]{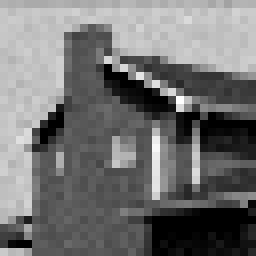}
	\includegraphics[width=0.3\linewidth]{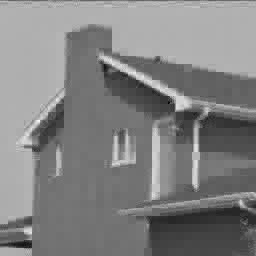}    

    \includegraphics[width=0.3\linewidth]{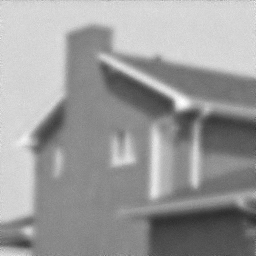}
	\includegraphics[width=0.3\linewidth]{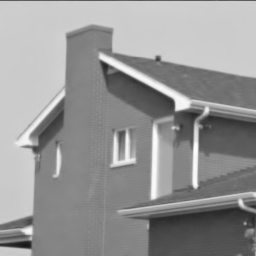}    
	\includegraphics[width=0.3\linewidth]{original_house.png}    
  	\caption{First row: Image reconstruction results for non-overlapping $8 \times 8$ patches from 32 Poisson-Gaussian CS measurements per patch with $\sigma = 200$, using (PG3) for $I = 10^8$ (left, RRMSE = $0.5$) and $I = 10^{10}$ (right, RRMSE = $0.07$). Second row (left and middle): Same as in the first row, but with overlapping patches and averaging in sliding window fashion: for $I = 10^8$ (left, RRMSE = $0.116$), $I = 10^{10}$ (right, RRMSE = $0.0326$). Second row (right): original image for reference.}
    \label{fig:imagerecon_PoissGauss}
\end{figure}

\textit{Image Reconstruction:}  Lastly, we ran an image-patch and image reconstruction experiment similar to the one described for Poisson noise. We simulated $N = 32$ measurements of the form $\*y_i \sim \textrm{Poisson}(\*\Phi_i \*x_i) + \*\eta_i$, for patch $\*x_i$ of $m = 64$ pixels. The $\sigma$ for $\*\eta_i$ was 200. The reconstruction was done independently patch-wise by solving (PG3) using $\*\Psi$ as the 2D-DCT basis and $\varepsilon = 2\sqrt{N}$. Results are presented on the $256 \times 256$ house image, for image-intensity $I \in \{10^6, 10^8, 10^{10}\}$ in Fig. \ref{fig:imagerecon_PoissGauss}. Due to the high $\sigma$ relative to the measurement values, the reconstruction failed at $I = 10^6$ and is not reported here, but improved for higher intensities. Compared to Fig. \ref{fig:imagerecon_Poiss}, the results in Fig. \ref{fig:imagerecon_PoissGauss} show higher RRMSE on non-overlapping blocks due to the presence of Gaussian noise. (The errors in both cases reduce upon sliding window averaging.) These experiments are evidence that our method works for compressible signals.

\section{Conclusion, Comparisons to Prior Art and Future Work}
\label{sec:discussion}
\textit{Contributions:} We have presented a convex implementable estimator for sparse/compressible signal reconstruction from CS measurements acquired by realistic sensing models, but corrupted by Poisson or Poisson-Gaussian noise. The estimator allows for statistically motivated and principled parameter tuning. To the best of our knowledge, there is no earlier work on analyzing Poisson CS using VSTs since the VSTs convert a problem with linear measurements to non-linear measurements \cite{Harmany2012}. We have demonstrated here, both theoretically as well as experimentally, that the non-linearity is actually not a problem, and that it does in fact have some advantages over the Poisson negative log-likelihood - namely more intuitive parameter tuning, besides Lipschitz continuity of the objective function and its derivative for $c \neq 0$. This is our first major contribution. Our second major contribution is the unification of analysis of Poisson CS and Poisson-Gaussian CS that our VST-based framework so readily allows for. Also ours is the first work to develop bounds for Poisson-Gaussian CS to the best of our knowledge. The extension of our method to Poisson-Gaussian noise also retains all the advantages of the method for Poisson noise.

\textit{Comparisons:} The previous work on Poisson CS in \cite{Raginsky2010,Jiang2015} applies to physically realizable sensing matrices but the theory there is developed only for computationally intractable estimators, and the latter work applies only to sparse (and not compressible) signals. The work in \cite{Rohban2016,Jiang2015_arxiv} applies to computationally tractable estimators (using the Poisson log-likelihood and the LASSO respectively), but does not explicitly address the important case of flux-preserving matrices. Recent work from \cite{Li2016_arxiv} applies to computationally tractable estimators, physical constraints and for sparse/compressible signals, but the estimator requires prior knowledge of a reasonable upper bound on signal sparsity, unlike our technique which has an easier choice of parameter during implementation. (In particular, the constraint $\|\boldsymbol{x}\|_1 = I$ was required only for the theoretical analysis and was not deemed necessary in the actual results. Imposition of this constraint is in fact not required even for the theoretical analysis if $\boldsymbol{\Phi}$ obeys the RIP). Also, besides our conference paper \cite{Garg2017}, our group has performed some other earlier work on Poisson CS for realistic matrices using a tractable estimator based on the Jensen-Shannon divergence (JSD) between $\boldsymbol{y}$ and $\boldsymbol{\Phi x}$ \cite{Patil2016_arxiv}. The work essentially makes use of the fact that the square-root of the JSD (SQJSD) is a metric, and that the SQJSD has values that scale as $o(\sqrt{N})$ but independent of $I$. In Table \ref{tab:comparisons}, we show comparisons of our work in this paper to six of the aforementioned, very recent techniques.

There exist other papers which provide performance guarantees for some variant of the LASSO for Poisson-related problems. For example, \cite{Rish2009} and \cite{Kakade2010} provide bounds using the RIP and maximum eigenvalue condition respectively. Necessary and sufficient conditions are derived for the sign consistency of the LASSO with the Poisson noise model in \cite{Jia2013}. Weighted/adaptive LASSO and group LASSO schemes with provable guarantees based on Poisson concentration inequalities have been proposed in \cite{Ivanoff2016,Jiang2015_arxiv}. The consistency of an $\ell_1$ regularized maximum likelihood (ML) estimator for compressive inversion is examined in \cite{YHLi2015} under the model $\lambda = \exp(-\boldsymbol{a}^t \boldsymbol{\theta})$ where $\boldsymbol{a}$ is a known vector, $\boldsymbol{\theta}$ is an unknown vector of sparse coefficients and $\lambda$ is the mean of the Poisson distribution. This work in fact shows that the regularization parameter is dependent on the signal sparsity, which is unknown in practice. Moreover, none of these techniques however explicitly deal with flux-preserving matrices. Also, the LASSO is not a probabilistic estimator in the Poisson case, as even a Gaussian approximation to the Poisson entails variances that are different for each measurement, and which are unknown during the estimation process. The LASSO has been extended to deal with non-linear problems in \cite{Plan2016,Yang2015_arxiv}, of which our technique is a special case (albeit with an additional non-negativity constraint). The technique in \cite{Yang2015_arxiv} derives error bounds on any stationary point of the objective function $\|\*y - f(\* \Phi \* x)\|^2 + \rho \|\* x\|_1$ for any differentiable monotonic function $f$ with bounded derivatives. At this point, we have not succeeded in adapting the technique from \cite{Yang2015_arxiv} to Poisson CS via the VST, because such an adaptation requires imposition of the additional necessary constraint $\* x \succeq \* 0$ while obtaining the stationary point of the objective function.

\textit{Future Work:} There are many directions for future work: (1) a derivation of lower bounds, (2) analysis of support recovery and prediction bounds $\|\boldsymbol{\Phi x} - \boldsymbol{\Phi {x^{\star}}}\|_2$, (3) analysis of the effect of clipping on Poisson-Gaussian CS measurements due to the limited dynamic range of sensors, (4) analysis using the original Poisson-Gaussian likelihood, as used in \cite{Chouzenoux2015} for deblurring, and (5) seeking an explanation for the good reconstruction results obtained even after ignoring the $\|\*x\|_1 = I$ constraint.

\begin{table}
\caption{Comparison of various methods analyzing performance bounds in Poisson and Poisson-Gaussian CS (Y = Yes, N = No)}
\begin{center}
  \begin{tabular}{| p{1.1 cm} | p{0.7 cm} | p {0.7 cm} | p {0.7 cm} | p {0.7 cm} | p {0.7 cm} | p {0.7 cm} | p{0.55 cm} |}
    \hline
    Feature & Our Method & \cite{Raginsky2010} & \cite{Jiang2015} & \cite{Rohban2016} & \cite{Jiang2015_arxiv} & \cite{Li2016_arxiv} & \cite{Patil2016_arxiv} \\
    \hline
	Tractable 
estimator & Y & N & N & Y & Y & Y & Y \\ \hline 
	Flux-preserving matrices & Y & Y & Y & N & N & Y & Y \\ \hline
	Sparse, compressible signals & Y & Y & Sparse only & Y & Y & Y & Y\\ \hline
	Parameters in estimator & None (or statistically motivated $\varepsilon$ ) & Y (regularization parameter) & Y (signal $\ell_0$ norm) & Y (signal $\ell_1$ norm) & Y (regularization parameter) & Y (signal $\ell_q$ norm, $q \leq 1$) & None (or statistically motivated $\varepsilon$) \\ \hline
	Lipschitz-continuity of objective function and its derivative & Y,Y (for $c \neq 0$) & N,N & N,N & N,N & Y,Y & N,N & N,N \\ \hline
	Maximum-likelihood based estimator & N & Y & Y & Y & N & Y & N \\ \hline
	Lower bounds derived & N & N & Y & Y & N & Y & N\\ \hline
	Extension to Poisson-Gaussian noise & Y & N & N & N & N & N & N\\ \hline
	Non-linear CS problem & Y (due to VST) & N & N & N & N & N & N\\ \hline
	  \end{tabular}
\label{tab:comparisons}
\end{center}
\end{table}

\section{Proofs}
\subsection{Proof of Theorem 1}
\label{subsec:thm1}
To prove theorem 1, we first begin by considering the case of a scalar $y \sim \textrm{Poisson}(\gamma)$ and generalize later to the case of measurement vectors. Define $f(y) \triangleq (\sqrt{y+c}-\sqrt{\gamma+c})^2$. Hence $f^{(1)}(y) = 1-\sqrt{\dfrac{\gamma+c}{y+c}}$, $f^{(2)}(y) = \dfrac{\sqrt{\gamma+c}(y+c)^{-1.5}}{2}$, and $f^{(3)}(y)=\dfrac{-3\sqrt{\gamma+c}(y+c)^{-2.5}}{4}$ where $f^{(k)}(y)$ denotes the $k^{th}$ derivative of $f(y)$ at $y$. Now, observe that $f(\gamma) = 0, f^{(1)}(\gamma) = 0$. Now $f(y) = f(\gamma) + \int_{\gamma}^{y} f^{(1)}(t) dt = \int_{\gamma}^{y} f^{(1)}(t) dt \leq (y-\gamma) f^{(1)}(y)$ since $f^{(1)}(y)$ is an increasing function of $y$. Similarly, we have $f^{(1)}(y) = f^{(1)}(\gamma) + \int_{\gamma}^{y} f^{(2)}(t) dt = \int_{\gamma}^{y} f^{(2)}(t) dt \leq (y-\gamma) f^{(2)}(\gamma)$ since $f^{(2)}(y)$ is a decreasing function. Combining this, we have 
\begin{equation}
f(y) \leq (y-\gamma) f^{(2)}(\gamma) = \dfrac{(y-\gamma)^2}{2(\gamma+c)}. 
\label{eq:fy}
\end{equation}
\\
Recall that $f(y)$ is a random variable. Taking expectation on both sides, we obtain
\begin{equation}
E[f(y)] \leq \dfrac{E[(y-\gamma)^2]}{2(\gamma+c)} \leq 0.5 \textrm{ as } E[(y-\gamma)^2] = \gamma.
\label{eq:E_fy}
\end{equation}
To obtain an upper bound on the variance of $f(y)$, we need a lower bound on $E[f(y)]$ since $\textrm{Var}(f(y)) = E[(f(y))^2] - (E[f(y)])^2$. To this, consider the following second order Taylor series expansion of $f(y)$ around $\gamma$ with a third-order Lagrange remainder term:
\begin{eqnarray}
f(y) = f(\gamma) + (y-\gamma)f^{(1)}(\gamma) + \frac{(y-\gamma)^2}{2!}f^{(2)}(\gamma) + \\ \nonumber
\frac{(y-\gamma)^2}{3!}f^{(3)}(z(y)),
\end{eqnarray} 
where $z(y) \in (\gamma,y)$ or $z(y) \in (y,\gamma)$. Using previous results for the derivatives, we have:
\begin{equation}
f(y) = \frac{(y-\gamma)^2}{4(\gamma+c)} -\frac{\sqrt{\gamma+c}(y-\gamma)^3}{8 (z(y)+c)^{2.5}}.
\end{equation} 
Taking expectation on both sides, we have
\begin{equation}
E[f(y)] = \frac{\gamma}{4(\gamma+c)} -\frac{\sqrt{\gamma+c}}{8} \sum_{y=0}^{\infty} (y-\gamma)^3 (z(y)+c)^{-2.5} e^{-\gamma} \gamma^y / y!.
\end{equation} 
Considering $\beta$ to be the largest integer less than or equal to $\gamma$, we can split the infinite summation in the equation above into two parts: one is a summation $K_1$ from $y=0$ to $y=\beta$, and the other is a summation $K_2$ from $y=\beta+1$ to $y=\infty$. In other words, we have
\begin{eqnarray}
K_1 = -\frac{\sqrt{\gamma+c}}{8} \sum_{y=0}^{\beta} (y-\gamma)^3 (z(y)+c)^{-2.5} e^{-\gamma} \gamma^y / y! \\ \nonumber
K_2 = -\frac{\sqrt{\gamma+c}}{8} \sum_{y=\beta+1}^{\infty} (y-\gamma)^3 (z(y)+c)^{-2.5} e^{-\gamma} \gamma^y / y!.
\end{eqnarray} 
To lower bound $E[f(y)]$ we seek a value of $z(y)$ which will minimize $K_1$ and a value of $z(y)$ which will maximize $K_2$. This is because $K_1$ is non-negative since $y \leq \gamma$ for terms in $K_1$, and $K_2$ is negative since $y > \gamma$ for terms in $K_2$. As $(z(y)+c)^{-2.5}$ is a decreasing function, we get $z(y) = \gamma$ in both cases. This yields
\begin{eqnarray}
E[f(y)] \geq \frac{\gamma}{4(\gamma+c)} -\frac{\sqrt{\gamma+c}}{8} (\gamma+c)^{-2.5} E[(y-\gamma)^3] \\
= \frac{\gamma}{4(\gamma+c)} - \frac{\gamma}{8(\gamma+c)^2}.
\end{eqnarray}
Here we have made use of the fact that $E[(y-\gamma)^3] = \gamma$ for a Poisson random variable $y$ with mean $\gamma$. As $f(y)$ is non-negative, we can write instead write 
\begin{equation}
E[f(y)] \geq \textrm{max}(0,\frac{\gamma}{4(\gamma+c)} - \frac{\gamma}{8(\gamma+c)^2}).
\end{equation}
Squaring both sides of Eqn. \ref{eq:fy} and taking expectation, we have
\begin{equation}
E[(f(y))^2] \leq \dfrac{E[(y-\gamma)^4]}{4(\gamma+c)^2} = \dfrac{\gamma(1+3\gamma)}{4(\gamma+c)^2},
\end{equation}
since $E[(y-\gamma)^4] = \gamma(1+3\gamma)$ for a Poisson random variable $y$ with mean $\gamma$. So we have
\begin{eqnarray}
\textrm{Var}[f(y)] = E[(f(y))^2] - (E[f(y)])^2 \\ 
\leq \dfrac{\gamma(1+3\gamma)}{4(\gamma+c)^2} - \textrm{max}(0,\frac{\gamma}{4(\gamma+c)} - \frac{\gamma}{8(\gamma+c)^2})^2 \\
\leq \dfrac{\gamma(1+3\gamma)}{4(\gamma+c)^2} \leq 3/4.
\label{eq:Var_fy}
\end{eqnarray}
The last inequality follows using L'Hospital's rules and using the fact that $\dfrac{\gamma(1+3\gamma)}{4(\gamma+c)^2}$ is a strictly increasing function of $\gamma$. We have so far derived upper bounds on the mean and variance of $f(y)$. Now we move to the case of a vector, i.e. to the case where $\boldsymbol{y}$ is a vector of $N$ measurements, where the $i^{th}$ measurement is given as $y_i \sim \textrm{Poisson}(\gamma_i)$ where $\gamma_i = (\boldsymbol{\Phi x})_i$. We also define $f_i(y_i) \triangleq (\sqrt{y_i+c}-\sqrt{\gamma_i+c})^2, f(\boldsymbol{y}) \triangleq \sum_{i=1}^N f_i(y_i), g(\boldsymbol{y}) \triangleq \sqrt{f(\boldsymbol{y})}$. Hence we have
$E[g(\boldsymbol{y})] = E[\sqrt{f(\boldsymbol{y})}] \leq \sqrt{E[f(\boldsymbol{y})]} \leq \sqrt{N/2}$ using Eqn. \ref{eq:E_fy}. This provs the first statement of Theorem 1.
\\
To derive a bound for the variance of $g(\boldsymbol{y})$, we proceed as follows. Define $\tilde{f}(\boldsymbol{y}) = f(\boldsymbol{y})/E[f(\boldsymbol{y})]$. Using the non-negativity of $\tilde{f}(\boldsymbol{y})$, we have 
\begin{equation}
\sqrt{\tilde{f}(\boldsymbol{y})} \geq 1 + (\tilde{f}(\boldsymbol{y})-1)/2  - (\tilde{f}(\boldsymbol{y})-1)^2/2.
\label{eq:sqrtf}
\end{equation}
To see why, consider that $l(h) \triangleq 3h-h^3 \leq 2$ for all $h \geq 0$ since $l(1) = 2$ and $l(h)$ is monotonically increasing in $[0,1]$ and monotonically decreasing in $[1,\infty)$. Putting $h = \sqrt{\tilde{f}}$ yields $3\sqrt{\tilde{f}}-\tilde{f}^{1.5} \leq 2 \rightarrow 3\tilde{f}-\tilde{f}^2 \leq 2 \sqrt{\tilde{f}}$ which after simple algebra yields Eqn. \ref{eq:sqrtf}. Taking expectation on both sides of Eqn. \ref{eq:sqrtf}, we have
\begin{equation}
E[\sqrt{\tilde{f}(\boldsymbol{y})}] \geq 1 - \textrm{Var}(\tilde{f}(\boldsymbol{y}))/2.
\end{equation}
Substituting the definition of $\tilde{f}(\boldsymbol{y})$, we have
\begin{equation}
E[g(\boldsymbol{y})] = E[\sqrt{f(\boldsymbol{y})}] \geq \sqrt{E[f(\boldsymbol{y})]}(1-\textrm{Var}(f(\boldsymbol{y}))/(2 (E[f(\boldsymbol{y})])^2).
\end{equation}
Since $\textrm{Var}(g(\boldsymbol{y})) = E[f(\boldsymbol{y})] - (E[g(\boldsymbol{y})])^2$, we have 
\begin{eqnarray}
\textrm{Var}(g(\boldsymbol{y})) \leq \dfrac{\textrm{Var}[f(\boldsymbol{y})]}{E[f(\boldsymbol{y})]} - \dfrac{(\textrm{Var}[f(\* y)])^2}{4 (E[f(\boldsymbol{y})])^3} \\ \nonumber
\leq \dfrac{\textrm{Var}[f(\boldsymbol{y})]}{E[f(\boldsymbol{y})]} =  \dfrac{\sum_{i=1}^N \textrm{Var}[f_i(y_i)]}{\sum_{i=1}^N E[f_i(y_i)]}.
\end{eqnarray}
Using the upper bound on $\textrm{Var}[f_i(y_i)]$ and the lower bound on $E[f_i(y_i)]$, we have the following bound on the variance:
\begin{equation}
\textrm{Var}(g(\boldsymbol{y})) \leq \dfrac{\sum_{i=1}^N \frac{3}{4} \frac{\gamma^2_i}{(\gamma_i+c)^2} + \frac{\gamma_i}{4(\gamma_i+c)^2}} {\sum_{i=1}^N \textrm{max}(0,\frac{\gamma_i}{4(\gamma_i+c)}-\frac{\gamma_i}{8(\gamma_i+c)^2})}.
\end{equation}
This proves the second statement of Theorem 1. For the third statement, observe that each term in the summation in the numerator is upper bounded by 1, leading to a numerator upper bound of $N$. Moreover one can show tha the term in the denominator is monotonically increasing for $\gamma_i \geq 1$ and hence is lower bounded by $\frac{2c+1}{8(c+1)^2}$. This proves the third statement, and the approximate value of $10.85$ can be obtained by using $c = \frac{3}{8}$.
\\
\\
In order to obtain a tail bound on $R(\* y, \*\Phi\*x)$ under the condition that $\*\Phi \*x \succeq \*1$, we can use Chebyshev's inequality to prove that $P(R(\* y, \*\Phi\*x) \leq \sqrt{N/2} + 3.29 \sqrt{N}) \geq 1-\frac{1}{N}$, since the variance of $R(\* y, \*\Phi\*x)$ is upper bounded by (approximately) $10.85$ when $\*\Phi \*x \succeq \*1$. This proves the fourth statement of the theorem.
\\
However, we show here that for large value of $N$, $R(\* y, \*\Phi\*x)$ is approximately Gaussian distributed which leads to tighter bounds and with an even higher probability: $P(R(\* y, \*\Phi\*x) \leq \sqrt{N/2} + \sqrt{3/4}\sqrt{N}) \geq 1-2e^{-N/2}$ using upper bounds on the mean and variance of $R(\* y, \*\Phi\*x)$. 
\\
By the CLT, we know that $P(\frac{f(\*y)-N\mu}{\sigma\sqrt{N}} \leq \alpha) \rightarrow \Phi_g(\alpha)$ as $N \rightarrow \infty$, where $\Phi_g$ is the CDF for $\mathcal{N}(0,1)$, and $\mu, \sigma$ are respectively the expected value and standard deviation of $f_i$. All the $f_i$ values have variances upper bounded by $3/4$ if $\*\Phi \*x \succeq \*1$. Due to the continuity of $\Phi_g$\footnote{inspired from \url{https://stats.stackexchange.com/questions/241504/central-limit-theorem-for-square-roots-of-\\sums-of-i-i-d-random-variables}}, we have $P(\frac{f(\*y)-N\mu}{\sigma\sqrt{N}} \leq \alpha + \frac{\alpha^2 \sigma^2}{4\mu\sigma\sqrt{N}}) \rightarrow \Phi_g(\alpha)$ as $N \rightarrow \infty$. Hence we have $P(f(\*y) \leq (\sqrt{N\mu}+\frac{\alpha \sigma}{2\sqrt{\mu}})^2) \rightarrow \Phi_g(\alpha)$ as $N \rightarrow \infty$, and taking square roots we get $P(\sqrt{f(\*y)} \leq (\sqrt{N\mu}+\frac{\alpha \sigma}{2\sqrt{\mu}})) \rightarrow \Phi_g(\alpha)$ as $N \rightarrow \infty$. By rearrangement, we obtain $P(\frac{\sqrt{f(\*y)}-\sqrt{N\mu}}{\sigma/(2\sqrt{\mu})} \leq \alpha) \rightarrow \Phi_g(\alpha)$ as $N \rightarrow \infty$. With this development and since $\mu \leq 1/2, \sigma^2 \leq 3/4$ from Eqns. \ref{eq:E_fy} and \ref{eq:Var_fy}, we can now invoke a Gaussian tail bound to establish that $P(R(\* y, \*\Phi\*x) \leq \sqrt{N/2} + \sqrt{3/4}\sqrt{N}) \geq 1-2e^{-N/2}$. Note that the Gaussian nature of $R(\* y, \*\Phi\*x)$ emerges from the CLT and is only an asymptotic result. However we consistently observe it to be approximately true even for small values of $N \sim 20$ as confirmed by a Kolmogorov-Smirnov test (see \cite{suppcode}). $\Box$
\vspace{-0.3in}
\subsection{Proof of Theorem 2}
\label{subsec:thm2}
We provide a sketch of the proof below, inspired from \cite{Candes2008}, but modified to suit our problem.
\begin{enumerate}
\item Define a vector $\boldsymbol{h} \triangleq \boldsymbol{\theta}-\boldsymbol{\theta^\star}$. Denote vector $\boldsymbol{h}_T$ to be equal to $\boldsymbol{h}$ only for index set $T$ and zero for other indices.  Let $T_0$ be the set containing $s$ largest absolute value indices of $\boldsymbol{h}$, $T_1$ be the set containing $s$ largest absolute value indices of $h_{T_0^c}$ and so on, where $T^c$ is the complement of the set $T$. Thus, vector $\boldsymbol{h}$ can be decomposed as the sum of $\boldsymbol{h}_{T0}$,$\boldsymbol{h}_{T_1}$,$\boldsymbol{h}_{T_2}$,...
\item Define $\boldsymbol{A} \triangleq \boldsymbol{\Phi \Psi}$. We have
\begin{align}
\|\boldsymbol{Ah}\|_2^2 
&= \|\boldsymbol{A}(\boldsymbol{\theta}-\boldsymbol{\theta^\star})\|_2^2 \nonumber
\\ &= \Sigma_{i=1}^N \Bigg( \bigg(\sqrt{(\boldsymbol{A \theta})_i+c} - \sqrt{(\boldsymbol{A\theta^\star})_i+c}\bigg)^2 \\ \nonumber
 &\bigg(\sqrt{(\boldsymbol{A\theta})_i+c} + \sqrt{(\boldsymbol{A\theta^\star})_i+c}\bigg)^2 \Bigg). 
\label{eq2}
\end{align}
\begin{enumerate}
\item Consider an upper bound of $\varepsilon$ on $\|\sqrt{\* y + c}-\sqrt{\* \Phi \* x + c}\|_2$. Later on, we shall assign a statistical meaning to $\varepsilon$ based on Theorem 1. By triangle inequality and the nature of the constraint in (P1), we have
\begin{align}
\|\sqrt{\boldsymbol{A\theta}+c}-\sqrt{\boldsymbol{A\theta^\star}+c}\|_2 &\leq \\ \nonumber
\|\sqrt{\boldsymbol{y}+c} - \sqrt{\boldsymbol{A\theta}+c}\|_2 + \\ \nonumber
\|\sqrt{\boldsymbol{y}+c} - \sqrt{\boldsymbol{A\theta^\star}+c}\|_2 &\leq 2\varepsilon. \label{eq3}
\end{align}
\item For scalars $v_1 \geq 0, v_2 \geq 0$, we have $(\sqrt{v_1} + \sqrt{v_2})^2 \leq 4 \textrm{max}(v_1,v_2)$. We also have $(\boldsymbol{A\theta})_i = (\boldsymbol{\Phi x})_i = \Sigma_j\Phi_{ij}x_j \leq \dfrac{\|\boldsymbol{x}\|_1}{N} = \dfrac{I}{N} $. Likewise $(\boldsymbol{A \theta^\star})_i \leq \dfrac{I}{N}$ as well, since $\|\boldsymbol{x^\star}\|_1 = I$. Hence $(\sqrt{(\boldsymbol{A \theta})_i+c} + \sqrt{(\boldsymbol{A \theta^\star})_i+c})^2 \leq 4(\dfrac{I}{N}+c)$.
\item Combining the earlier two results with Eqn. \ref{eq2}, we have $\|\boldsymbol{Ah}\|_2 \leq 4\varepsilon \sqrt{\dfrac{I}{N}+c}$.
\end{enumerate}
\item To prove the bound on $\|\mathbf{h}_{(T_0 \cup T_1)^c}\|_2$, we follow steps similar to \cite{Candes2008} to obtain 
\begin{equation}\label{eq9}
\|\mathbf{h}_{(T_0 \cup T_1)^c}\|_2 \leq \|\mathbf{h}_{{(T_0)}}\|_2 + 2s^{-1/2} \|\boldsymbol{\theta}-\boldsymbol{\theta_s}\|_1.
\end{equation}
\item To prove error bounds on $\|\mathbf{h}_{(T_0 \cup T_1)}\|_2$, we adopt the following steps.
\begin{enumerate}
\item Given the construction for $\boldsymbol{\Phi}$ in Eqn. \ref{eq:Phi}, we have
\begin{align}
\boldsymbol{\Phi \Psi(\theta-\theta^\star)} &= \frac{1}{2\sqrt{N}} \boldsymbol{\widetilde{\Phi} \Psi}( \boldsymbol{\theta}-\boldsymbol{\theta^\star}) + \nonumber
\\ & (\|\boldsymbol{\Psi\theta}\|_1-\|\boldsymbol{\Psi\theta^\star}\|_1) \nonumber
\\ &= \frac{1}{2\sqrt{N}} \boldsymbol{\widetilde{\Phi} \Psi}( \boldsymbol{\theta}-\boldsymbol{\theta^\star})
\end{align}
since we know that $\|\boldsymbol{\Psi\theta}\|_1 = \|\boldsymbol{\Psi\theta^\star}\|_1 = I$.
Defining $\boldsymbol{B} \triangleq \boldsymbol{\widetilde{\Phi}\Psi}$, we get 
\begin{align}
\|\boldsymbol{Bh}\|_2 &= 2 \sqrt{N}\|\boldsymbol{Ah}\|_2 \leq 8\varepsilon\sqrt{I+cN}.
\end{align}
\item Following steps in \cite{Candes2008} using the RIP and the Cauchy-Schwarz inequality, we can prove that
\begin{align}
\|\boldsymbol{h}_{T_0 \cup T_1}\|_2 &\leq C'\varepsilon\hspace{0.1cm}\sqrt[]{I+cN} + C''s^{-1/2}\|\boldsymbol{\theta}-\boldsymbol{\theta^\star}\|_1
\end{align} 
where $C' \triangleq \frac{2\hspace{0.1cm}\sqrt{1+\delta_{2s}}}{1-\delta_{2s}(\sqrt{2}+1)}$ and $C'' \triangleq \frac{2\hspace{0.1cm}\sqrt{2} \delta_{2s}}{1-\delta_{2s}(\sqrt{2}+1)}$.
\end{enumerate}
\item Combining the bounds on $\|\boldsymbol{h}_{T_0 \cup T_1}\|_2$ and $\|\boldsymbol{h}_{{T_0 \cup T_1}^c}\|_2$, we have
\begin{equation}
\|\boldsymbol{h}\|_2 \leq C_1\varepsilon\hspace{0.1cm}\sqrt{I+cN} + C_2\sqrt{2} \|\boldsymbol{\theta}-\boldsymbol{\theta_s}\|_1
\end{equation}
where $C_1 \triangleq 2C'$ and $C_2 \triangleq 2+2C''$.
\end{enumerate}
Finally, we divide by $I$ to obtain upper RRE bounds:
\begin{equation}
\frac{\|\boldsymbol{\theta}-\boldsymbol{\theta^\star}\|_2}{I} \leq C_1\varepsilon \sqrt{\frac{1}{I}+\frac{cN}{I^2}} + \frac{C_2s^{-\frac{1}{2}}\|\boldsymbol{\theta}-\boldsymbol{\theta_s}\|_1}{I}.
\end{equation}
%Based on Theorem 1, we see that the expected value of $\varepsilon$ is $\sqrt{N/2}$. Substituting this in the above equation proves the first part of Theorem 2. 
Using Theorem 1, we see that $\varepsilon \leq \sqrt{N}(3.29/\kappa+1/\sqrt{2})$ with a probability of $1-\kappa^2/N$ for any $\kappa > 0$. This proves Theorem 2. Note that both this bound makes appropriate use of the fact that $\* y$ is Poisson distributed. $\Box$

\subsection{Proof of Theorem 3}
\label{subsec:thm3}
The proof of this theorem is very similar to that of Theorem 1, so we mention only the points of difference. First, right through the proof, the constant $c$ is replaced by $d \triangleq c + \sigma^2$. Moreover for Poisson-Gaussian noise where the Gaussian component is signal-independent, we have $E[(y-\gamma)^2] = \gamma + \sigma^2, E[(y-\gamma)^3]=\gamma, E[(y-\gamma)^4]=\gamma(1+3\gamma)+\sigma^4$. Despite these changes, the upper bound for $E[f(y)]$ from Eqn. \ref{eq:E_fy} remains unchanged (and so does the lower bound for $E[f(\* y)]$). The upper bound for the variance of $f(y)$ from Eqn. \ref{eq:Var_fy} becomes $\textrm{Var}[f(y)] \leq \dfrac{\gamma(1+3\gamma)+\sigma^4}{4(\gamma+d)^2} \leq 3/4+1/4=1$. This step is again similar to that in Theorem 1, except that we have an added term $\dfrac{\sigma^4}{4(\gamma+d)^2}$ which is upper bounded by $1/4$. Following similar steps, the final upper bound for the variance of $g(\* y)$ is given by:
\begin{equation}
\textrm{Var}(g(\boldsymbol{y})) \leq \dfrac{\sum_{i=1}^N \frac{\gamma_i(1+3\gamma_i)+\sigma^4}{(\gamma_i+d)^2}} {\sum_{i=1}^N \textrm{max}(0,\frac{\gamma_i}{4(\gamma_i+d)}-\frac{\gamma_i}{8(\gamma_i+d)^2})}.
\end{equation}
The third statement of the theorem can also be easily derived using similar arguments, and these bounds can be approximately refined via the CLT to yield $P(R_d(\* y, \*\Phi\*x) \leq \sqrt{N/2} + \sqrt{N}) \geq 1-2e^{-N/2}$. 

\bibliographystyle{IEEEtran}
\bibliography{refs}

%\begin{thebibliography}{1}
  %0.5em minus 0.4em\relax Harlow, England: Addison-Wesley, 1999.
%\end{thebibliography}


%\end{IEEEbiography}


%\newpage



%\vfill


\end{document}